%% file: DAPR-2017-01-PAPER.tex
\pdfoutput=1
\newcommand*{\ATLASLATEXPATH}{latex/}
\documentclass[cernpreprint, atlasdraft=false, texlive=2016, UKenglish]{\ATLASLATEXPATH atlasdoc}

\usepackage[backend=bibtex]{\ATLASLATEXPATH atlaspackage}
\usepackage{\ATLASLATEXPATH atlasphysics}
\usepackage{\ATLASLATEXPATH atlasbiblatex}
\usepackage{\ATLASLATEXPATH atlascontribute}

\usepackage{\ATLASLATEXPATH atlasphysics}

\graphicspath{{logos/}{figures/}}

\usepackage{DAPR-2017-01-defs}

\input{DAPR-2017-01-metadata}

\hypersetup{pdftitle={ATLAS document},pdfauthor={The ATLAS Collaboration}}

\begin{document}

\maketitle

\hyphenation{}

\section{Introduction}
Proton losses in the Large Hadron Collider (LHC) ring upstream\footnote{Upstream and downstream are 
defined relative to the beam direction.} of the ATLAS experiment\,\cite{atlas}, 
due to interactions with either residual gas in the beam pipe (beam--gas scattering) 
or with machine elements such as collimators, result in beam-induced background (BIB).
Although the rates are negligible compared to particle debris 
from almost 10$^{9}$ proton--proton ($pp$) collisions per second, 
BIB has particular features that render it potentially problematic: 
it is characterised by particles almost parallel to the beam line,
which can produce elongated clusters with large energy deposition in the innermost tracking detectors
based on silicon pixel technology. 
At high rates, these abnormally large clusters can
affect data-taking efficiency\,\cite{THOMPSON2011133}.
Furthermore, a potential background for physics analyses arises from high-energy muons, originating mostly from
pion and kaon decay in the hadronic showers induced by beam losses.
These muons can deposit large amounts of energy in calorimeters through radiative processes. 
Such energy depositions, which are not associated with a hard scattering at the interaction point (IP), can 
be reconstructed as {\em fake jets} leading to missing transverse momentum if overlaid with a collision event.
Especially in searches for some exotic physics processes\,\cite{monojet2011,rhadron,longlivedpair2015,darkmatter}, fake jets
represent a non-negligible background that must be well controlled and subtracted.

Although BIB has had no detrimental effects on ATLAS operation so far, the continuous striving for
better LHC luminosity performance might change this situation in the future. A thorough understanding of 
the sources and nature of BIB, which is crucial when planning upgrades to the LHC, can only be achieved
by a combination of measurements and simulations. A validation of the latter is the main purpose of this work.

A lot of experience with BIB was gained at the Tevatron and HERA colliders\,\cite{tevatronref, herabackgrounds, ws2008}. 
The first simulation predictions for BIB at the LHC were presented more than 20 years ago\,\cite{nima381:531} and
have been refined several times thereafter\,\cite{nikolai1, nikolai2, lhcb-ieee, nima729:825}.
Throughout the LHC operation, BIB is routinely monitored and analysed by 
ATLAS\,\cite{backgroundpaper2011,backgroundpaper2012}. In this paper, comparisons 
between detailed simulations using the \fluka Monte Carlo (MC) simulation package\,\cite{flukaref1, flukaref2} and 
measurements\,\cite{backgroundpaper2012} of BIB during the 2012 LHC run, with a proton beam energy of 4\,\TeV, are presented.

\section{The LHC accelerator and the ATLAS experiment}
The LHC accelerator and the ATLAS experiment are described in detail in 
Refs.\,\cite{lhc1} and \cite{atlas} respectively.
Only a summary, focused on aspects relevant to the studies and simulations of BIB, is given here.  

\subsection{The LHC}
\begin{figure}[t]
\centering
\includegraphics[width=0.7\textwidth]{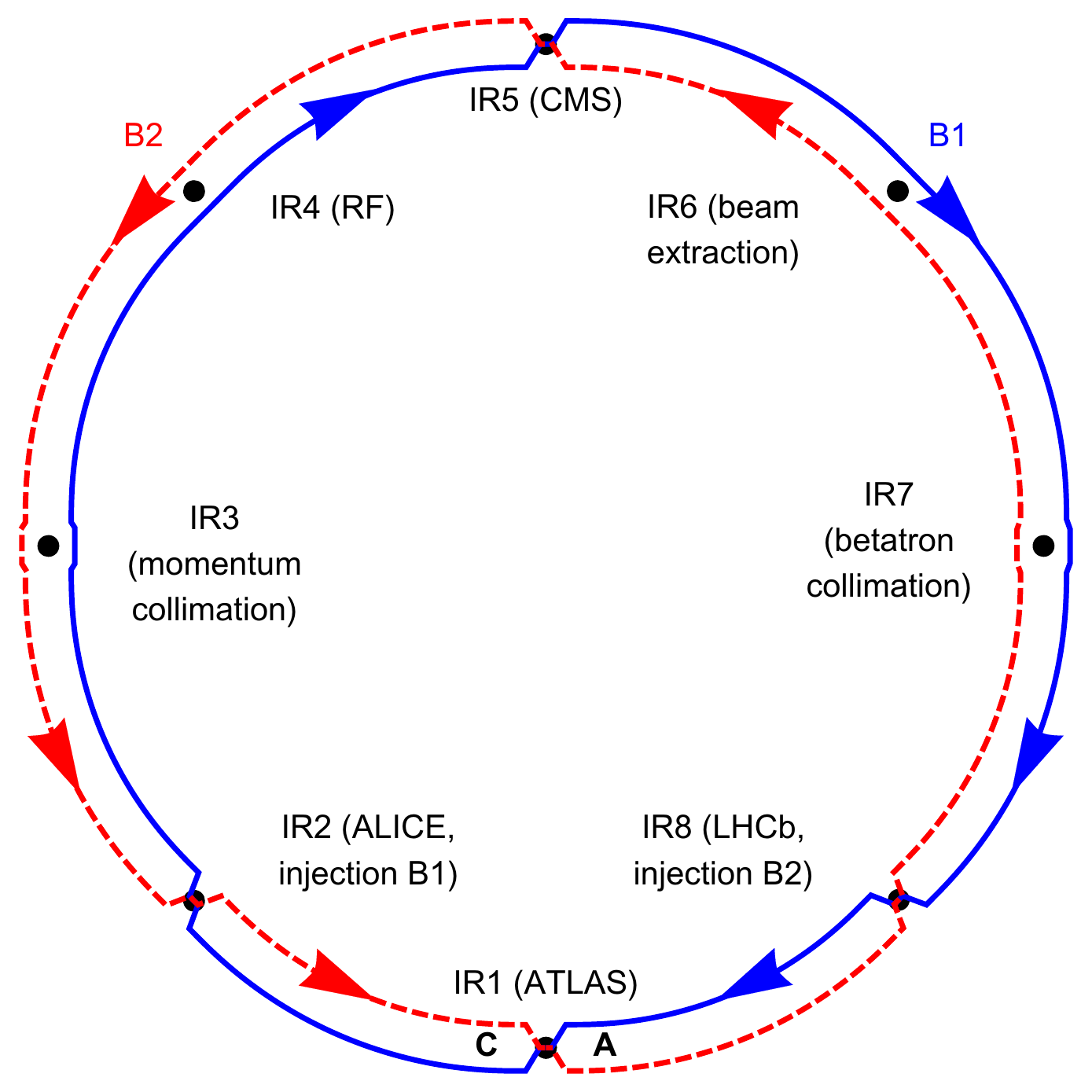}
\caption{The general layout of the LHC\,\protect\cite{lhc1}, showing the eight interaction 
regions. The counter-circulating beams are shown schematically, i.e. their separation is not
to scale. The ATLAS convention of labelling sides by `A' and `C' is indicated.
The figure is adapted from Ref.\,\protect\cite{nim848:19}.}
\label{fig:LHC}
\end{figure}

The LHC, shown schematically in Figure\,\ref{fig:LHC}, consists of eight arcs 
that are joined by long straight sections (LSSs) of $\sim$500\,m length. 
In the middle of each LSS there is an interaction region (IR); 
the ATLAS experiment is situated in IR1.
The LHC beam-cleaning equipment is located in IR3 and IR7, 
for momentum and betatron cleaning respectively. 
The principal performance parameters of LHC operation in 2012 are listed in Table\,\ref{tabLHC}. 

\begin{table}[t]
\renewcommand{\arraystretch}{1.3}
\caption{LHC parameters during operation as a $pp$ collider in the second half of 2012. The parameter $\beta^{*}$ refers 
to the value of the optical $\beta$-function at the collision point.}
\label{tabLHC}
\centering
\begin{tabular}{l|r}
\hline
Parameter &  Value \\
\hline
Beam energy [\TeV]                             & 4.0                 \\ 
Protons per bunch [10$^{11}$]                  & $\sim$1.5           \\
Number of bunches per beam                     & 1374\phantom{.0}    \\
Bunch spacing [ns]                             & 50\phantom{.0}      \\                    
Vertical crossing angle in IR1 [$\mu$rad]      & 145.0               \\
$\beta^{*}$ in IR1 [m]                         & 0.6                 \\
\hline
\end{tabular}
\end{table}

\begin{figure}[t]
\centering
\includegraphics[width=0.9\textwidth]{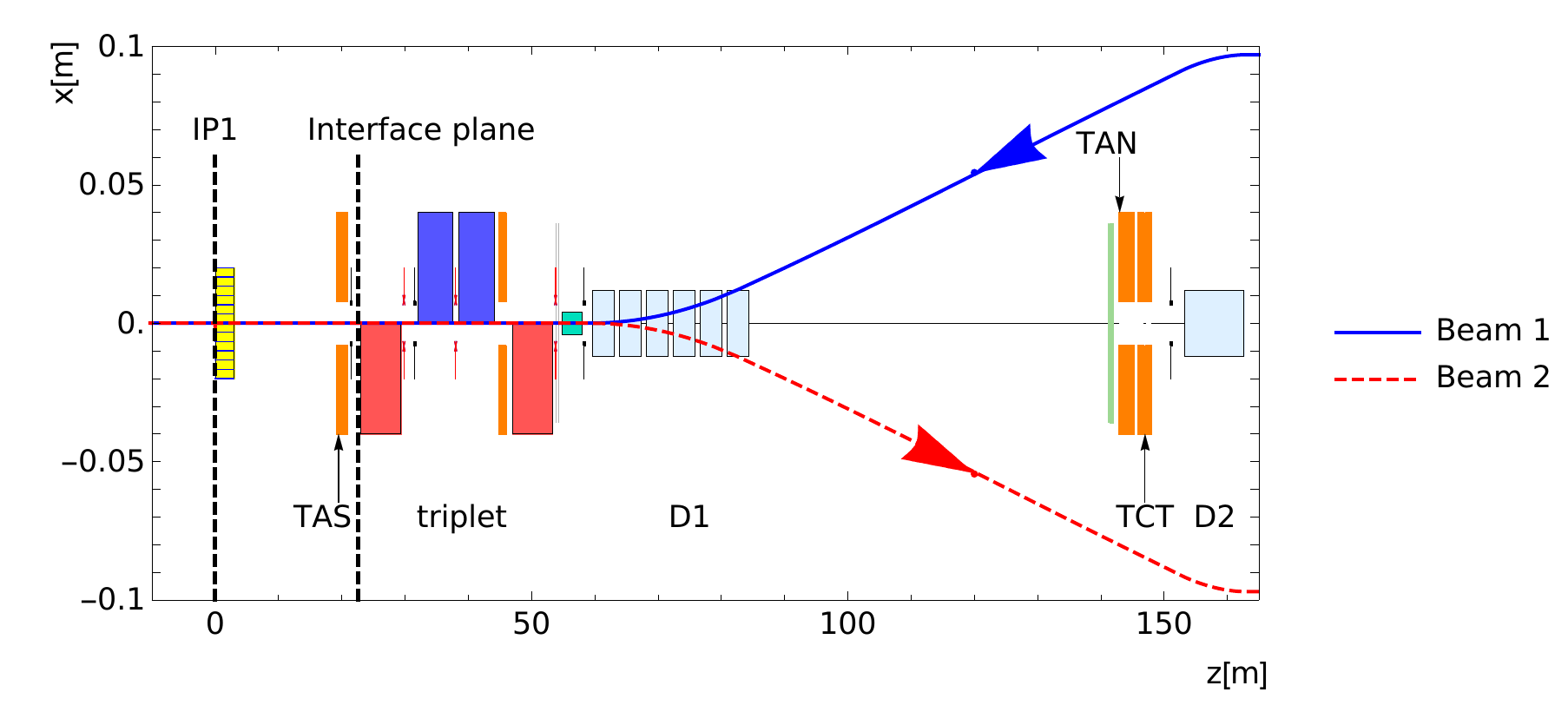}
\caption{Layout of the IR1 region showing the $z$-location of LHC beam-line elements 
and schematic beam trajectories. The $x$-coordinates refer only to the positions of the beams, not to the
beam-line elements. The beams are
separated by the D1 magnet and recombined into parallel trajectories by the D2 magnet. 
The tertiary collimator (TCT) is only on the incoming beam, just before the neutral particle absorber (TAN).
The sense of focusing of the four triplet elements is indicated by the colour of the boxes (red = vertical, 
blue = horizontal) for the incoming beam. The interface plane is explained in
Section\,\protect\ref{sect:simulations}.
} 
\label{fig:schematic_IR1}
\end{figure}

A schematic layout of IR1, up to 165\,m from the interaction point (IP, at $z=0$), is shown in Figure\,\ref{fig:schematic_IR1}, 
where the separation of the two counter-rotating beams is illustrated. 
Copper absorbers (TAS), which protect the superconducting inner triplets from collision debris, are
located between $|z|\!=\,$19\,m and $|z|\!=\,$20.8\,m and have an aperture of $r\!=\,$17\,mm.
The final focus is provided by the quadrupoles of the inner triplets on each side of the IP, 
between $|z|\!=\,$23\,m and $|z|\!=\,$54\,m.
The beam trajectories are separated at $|z|\!\approx\,$70\,m inside the separation dipoles D1 
and recombined in dipoles D2 at $|z|\!\approx\,$160\,m, which bring the two beams into parallel 
trajectories at a distance of 194\,mm from each other.
The dipole D2 is superconducting and is protected by the neutral particle absorber (TAN), which
intercepts energetic neutrons and photons emitted from the IP at very small angles. 

The 400.79\,MHz frequency of the LHC radio-frequency (RF) system and the revolution 
time of 88.9244\,$\mu$s form 35640 buckets that can be filled with particles.
In the 2012 LHC run, every 20$^{th}$ bucket was filled giving a bunch spacing of 50\,ns.
In order to facilitate monitoring of BIB, a few (typically six per beam in 2012) unpaired bunches 
are included in each LHC bunch pattern.
Having no counterpart in the other beam to collide with,
these bunches provide the LHC experiments with a rather clean measurement of BIB.

\subsection{The ATLAS experiment}
\label{sect:atlas}

The ATLAS experiment is one of the two general-purpose detectors at the LHC. 
With a length of 46\,m and a diameter of 25\,m, 
it is optimised to study proton--proton collisions at the highest available energies and luminosities.

In this study, the right-handed ATLAS coordinate system is used. The origin is at the nominal IP
and the azimuthal angle $\phi$ is measured relative to the $x$-axis, 
which points towards the centre of the LHC ring. 
Side A of ATLAS is defined as the side of the incoming, clockwise, LHC beam-1 
while the side of the incoming beam-2 is labelled C, as illustrated in Figure\,\ref{fig:LHC}. 
The $z$-axis points from C to A, i.e.\ along the beam-2 direction.
The pseudorapidity is given by $\eta\!=\!-\ln\tan(\theta/2)$, where $\theta$ is the polar angle  
relative to the $z$-axis. The transverse momentum is defined as $\pt=p\sin\theta$, where $p$ is 
obtained from the energy deposits in the calorimeters, assuming them to be massless.

ATLAS includes a dedicated beam conditions monitor (BCM)\,\cite{bcm} for beam background measurements.
The BCM consists of four small diamond modules 
on each side of the IP, at $z = \pm 1.84$\,m,
at a mean radial distance of $r\!=\!55$\,mm ($|\eta|\!\approx\,$4.2) from the beam line.
The modules are arranged in a cross: two in the horizontal and two in the vertical plane. 
Each module has two back-to-back sensors with an active area of 8$\times$8\,mm${^2}$ 
and a total thickness of 1\,mm. 

The inner detector\,\cite{innerDet} is subdivided into a pixel detector immediately outside the beam
pipe, a silicon-strip tracker and an outer transition-radiation tracker. These are inside a solenoid, 
which produces a 2\,T magnetic field along the $z$-axis.
The inner detector is used to determine the momentum of charged particles 
in the pseudorapidity range $|\eta|\!<\,$2.5.

The calorimeter system, which measures the energy of the particles, includes
a high-granularity liquid-argon (LAr) electromagnetic barrel calorimeter 
with lead as absorber; it has a half-length of  $\sim$3\,m and extends radially from $r\!=\,$1.5\,m to 2.0\,m,  
thus covering pseudorapidities up to $|\eta|\!=\,$1.5.
Between $r\!=\,$2.3\,m and 4.3\,m a scintillator-tile hadronic barrel calorimeter (Tile) 
with steel as absorber and $\sim$6\,m half-length covers pseudorapidities up to $|\eta|\!=\,$1.7.
The calorimeter system is extended, up to $|\eta|\!=\,$3.2, by electromagnetic and hadronic endcaps 
based on LAr technology. These have lengths, along $z$, of 0.6\,m and 1.8\,m respectively. 

The calorimeters are surrounded by a muon spectrometer based on three large air-core superconducting toroidal magnets
with eight coils each: one barrel toroid and two endcap toroids positioned inside the barrel at the ends of the central 
solenoid.

\section{Beam-induced background}
\label{sect:bib}
\begin{figure}[t]
\centering
\includegraphics[width=0.98\textwidth]{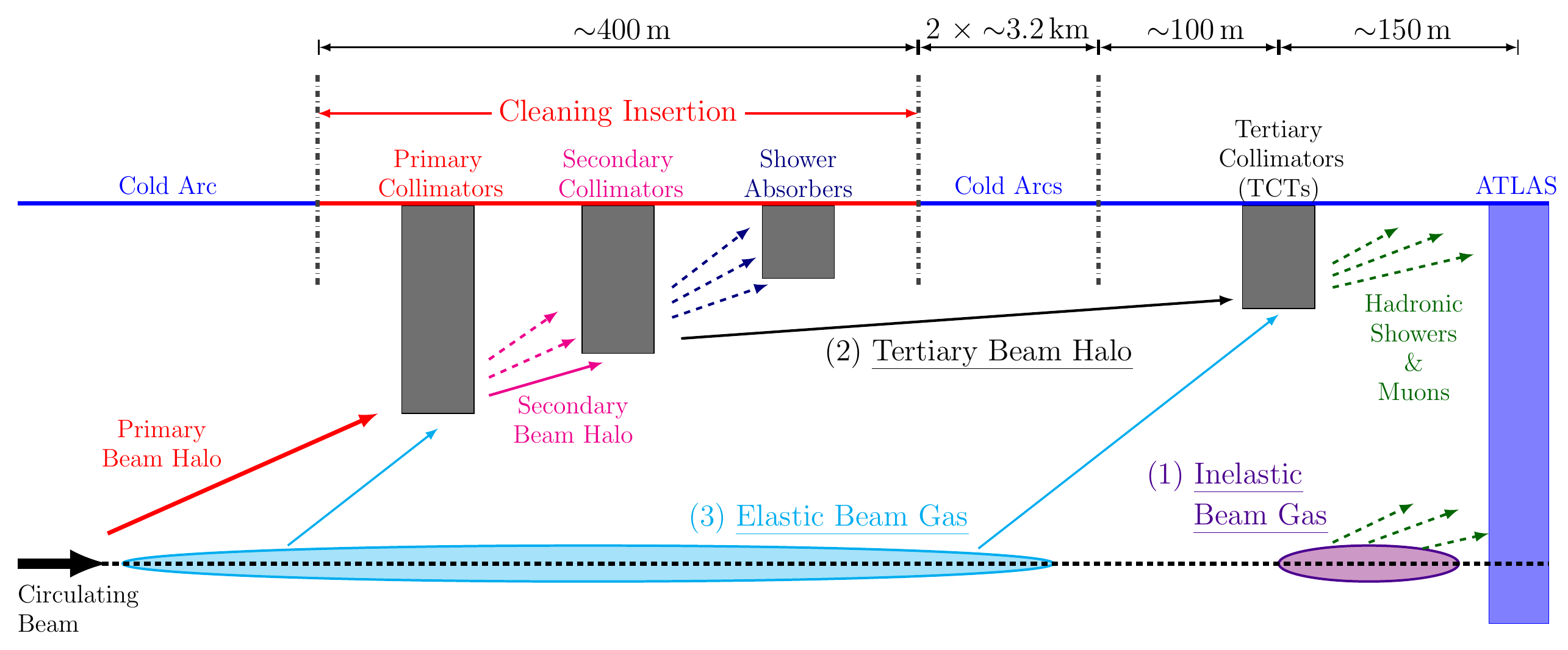}
\caption{Schematic illustration of the three sources of BIB reaching ATLAS:
(1) nearby inelastic beam--gas collisions, (2) tertiary beam halo losses on the TCT and (3) protons deflected
by elastic beam--gas collisions and hitting the TCT. The cleaning insertions are 6.7\,km away from ATLAS and the elastic beam--gas 
events are distributed around the entire accelerator ring. The distance from the beam to the collimators is a few millimetres.
}
\label{fig:schematic-Background}
\end{figure}

Beam induced background originates from three different beam-loss processes, which are illustrated 
in Figure\,\ref{fig:schematic-Background} and detailed below.

Inelastic proton interactions with residual gas inside the beam pipe (labelled 1 in Figure\,\ref{fig:schematic-Background}), 
in the vicinity of the IP, constitute the dominant source of BIB in ATLAS. Hadronic and electromagnetic showers, but in 
particular high-energy muons produced by these interactions, can enter ATLAS and be detected by 
the BIB monitoring system. It was shown in previous studies\,\cite{nima381:531} that inelastic
beam--gas collisions up to distances of $\sim$500\,m from the IP contribute to the background.

A small fraction of BIB arises from beam halo, which is continuously repopulated by scattering of 
particles from the beam due to various processes such as elastic collisions at the experiments and with
residual gas, noise on the RF system and feedback, intrabeam scattering, resonances and instabilities.
The superconducting magnets of the LHC require very efficient halo-cleaning, which is realised by a multi-step 
cleaning system\,\cite{RodCol}. 
The primary and secondary collimators of the cleaning insertions in IR3 and IR7 intercept most of the off-momentum and
betatron halo. A small fraction of the protons escape these insertions and constitute
the tertiary halo (labelled 2 in Figure\,\ref{fig:schematic-Background}) which is intercepted by the tertiary 
collimators (TCTs), located at distances of $z\!\approx\,$150\,m from each experimental IP.
Protons impinging on the TCTs can also originate from elastic beam--gas interactions (labelled 3 in Figure\,\ref{fig:schematic-Background}), which 
deflect protons out of the beam, around the whole accelerator ring.
The losses on the TCTs create showers, which can propagate all the way to the IP. 
Dedicated tests\,\cite{ATL-DAPR-PUB-2017-001} during 2015 and 2016 showed that, in normal physics conditions, 
total losses on the TCTs contribute of the order of only 1\% to the total BIB seen in ATLAS.
Thus they are not considered in this paper.

The rate of beam--gas interactions is proportional to the residual gas pressure and the beam intensity.
The latter is a property of the beams and is measured by the LHC with percent-level accuracy, but
the pressure and molecular composition of the residual gas varies as a function of position around the accelerator. 
After pumping down of the LHC beam vacuum, a small amount of gas remains stuck on the beam-pipe surface. These gas
molecules can be desorbed by synchrotron radiation or charged particles hitting the beam-pipe walls. The rate of outgassing
depends on the intensity of the radiation and therefore the dynamical pressure depends on beam intensity and energy.
In addition, the surface characteristics and temperature have a large influence on the residual pressure. 
The residual gas consists of H$_2$, CH$_{4}$, CO$_{2}$ and CO. Their relative fractions
depend on local temperature, radiation load and surface characteristics of the beam pipe. 
In cryogenic sectors the gas condenses on the cold walls, but is relatively easily released by irradiation. Almost all room-temperature
sectors of the LHC beam pipe are coated with a non-evaporable getter material\,\cite{neg-benvenuti}, which provides distributed 
pumping along the beam line for all common gases except CH$_4$. Therefore, methane is the dominant gas species in room-temperature sections, 
including the D1 dipole. 
Inside cryogenic magnets where the cold bore is at 1.9\,K, notably those of the inner triplet and the LHC arc, all gases except hydrogen 
stick relatively firmly on the surface, so the dominant gas is H$_2$.
The magnets in the LSS, from D2 to the arc, are operated at 4.5\,K. At this temperature all gases are more easily desorbed and 
here CO$_2$ is the most abundant gas species.
Vacuum pumps produce local minima in the pressure and corresponding gradients which result in gas diffusion from 
sections with higher pressure towards the pumps. 

The room-temperature sections of the LHC are equipped with vacuum gauges, but between these measurement points the pressure
has to be obtained from simulations. The simulation models are based on theory and laboratory measurements of
desorption rates and gas composition\,\cite{pMapSim}.
The amount of gas on the surfaces depends on the beam-conditioning history: when gas is desorbed and pumped out during 
beam operation the rate of outgassing slowly goes down. The state of surface conditioning, at any given time, has to be empirically 
estimated based on prior experience. 
Thus, the simulations depend on the local characteristics and temperature of the beam pipe, local pumping
speeds, beam intensity and the estimated effects of the beam-conditioning history. The overall uncertainty in the
local pressure due to knowledge of these parameters, especially of the state of surface conditioning, is estimated to be a 
factor of $\sim$3. 
Since both the surface characteristics and the intensity of radiation vary as a function of position, 
this uncertainty is not a global scale factor of the entire pressure distribution; it is possible that the pressure is underestimated in 
some regions and overestimated in others.

\section{Background monitoring methods}
The rates of BIB are measured by the BCM and the 
calorimeters, which are described in Section\,\ref{sect:atlas}. They both provide low-level
trigger signals which can be used for real-time background monitoring, and also record the data for detailed offline 
analysis. Only the unpaired bunches are used for monitoring and analysis of BIB. For inelastic beam--gas background
these can be assumed to be perfectly representative of colliding bunches.
								     
\subsection{BCM background rates}

\begin{figure}[t]
\centering
\includegraphics[width=0.8\textwidth]{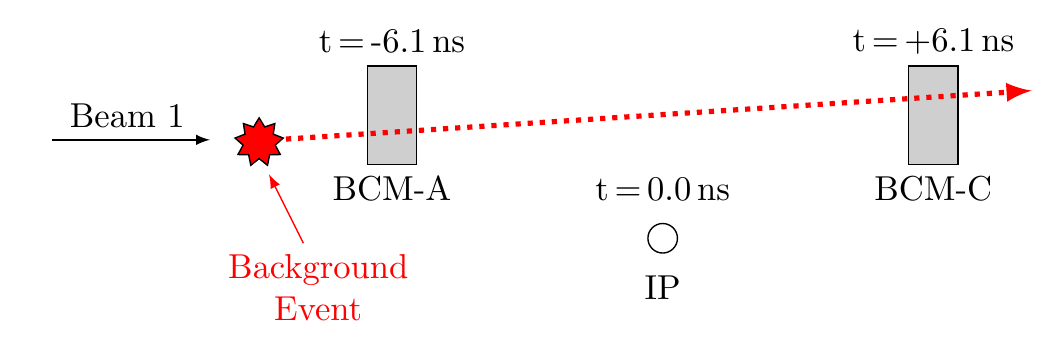}
\caption{Illustration of the BCM background trigger signature for beam-1.
The dotted line represents the trajectory of one particle 
hitting the upstream and downstream BCM modules.
For beam-1 the early hit is on side A and the in-time hit on side C. For beam-2 the
direction is reversed. The trigger can also be fired by two different particles, the only 
requirement being that any of the four modules on one side of the IP has an early hit and any of
those on the opposite side has an in-time hit.}
\label{fig:schematic_BCM_Trigger}
\end{figure}

Hits in the BCM modules are counted above a threshold of 250\,\keV, which corresponds to roughly 40\% of the 
energy deposition of a minimum-ionising particle in 1\,mm of diamond.
Particles from beam losses reach upstream BCM detectors 6.1\,ns before the nominal collision time, i.e.\
the passage of the bunch at the IP at $t=0$, and produce {\em early} hits. Both the BIB and
collision products from the IP produce {\em in-time} hits in the downstream detectors at $t = +6.1$\,ns.

A BCM background trigger signature, illustrated in Figure\,\ref{fig:schematic_BCM_Trigger},  
consists of an early hit in any module on the upstream side and an in-time hit in any module on the downstream side. 
The time windows of the background trigger are 5.46\,ns wide and nominally centred at $t=\pm 6.25$\,ns.
The BCM has sub-nanosecond time resolution and the nominal centre of the trigger window is aligned with 
the LHC collision time to an accuracy better than 2\,ns.
 
Due to the built-in direction requirement, the BCM background trigger is able to distinguish which beam 
the background originates from. In 2012 a single BCM background trigger, which fired on events in either 
direction, was used to collect events for the offline analysis.

\subsection{Fake jets in calorimeters}
\label{sec:fakejets}
The barrel and endcap calorimeters have nanosecond time resolution and contribute to a 
jet trigger with a \pt threshold of 10\,\GeV\ at the electromagnetic scale, which is used to select fake-jet 
candidates induced by BIB in unpaired bunches. 
The jets are reconstructed with the anti-$k_t$ jet algorithm\,\cite{Cacciari:2008gp}
with radius parameter $R$ = 0.4 using the {\textsc FastJet} software package\,\cite{Cacciari2012}. The inputs to this algorithm are 
topologically connected clusters of calorimeter cells\,\cite{jetCorRef}, seeded by cells with an energy at least four standard 
deviations above the measured noise. These topological clusters are calibrated at the electromagnetic scale. 
The reconstructed jets are corrected for contributions from additional $pp$ interactions in the same and neighbouring bunch
crossings as described in Ref.\,\cite{jetCorRef}. 
In order to suppress instrumental backgrounds, standard data-quality requirements are imposed\,\cite{Aad:2014una}. 
Data from periods affected by calorimeter noise bursts are excluded from the analysis.

\section{Simulation framework}
\label{sect:simulations}
\begin{figure}[t]
\centering
\includegraphics[width=0.98\textwidth]{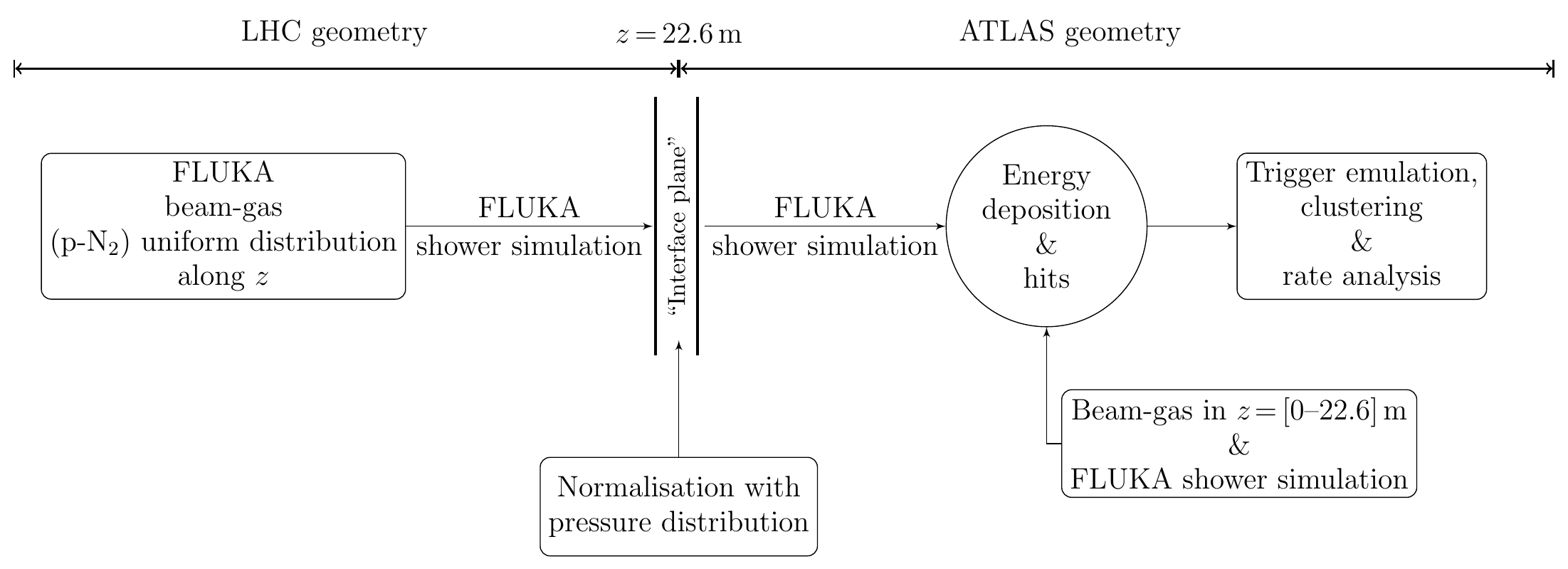}
\caption{Flowchart of the two-step process for simulation of beam--gas events. Particles crossing the
interface plane are stored in a file, and normalised using the appropriate pressure distribution
before being injected into the ATLAS simulation. The trigger emulation, clustering and rate analysis
are performed on the custom \fluka output.} 
\label{fig:simFlow}
\end{figure}

The simulation of the inelastic beam--gas events was performed with \fluka
using a two-step approach -- a method first introduced in Ref.\,\cite{nima381:531} and  illustrated in Figure\,\ref{fig:simFlow}. 
The advantage of dividing the simulation into accelerator- and detector-specific parts is that it leaves more flexibility in 
the choice of simulation tools. This approach also saves computational resources since the results of the first step,
simulation of particle transport and showering in the accelerator structures, can be used for several studies of 
the impact on the ATLAS detector.

The first step is discussed in detail in Ref.\,\cite{regina}:
beam--gas events with a uniform distribution in a $z$-range from 22.6\,m to 546.6\,m 
were generated with \fluka as inelastic $p$--N$_2$ interactions. Although the residual gas composition varies along the
ring and H$_2$ is most abundant in 1.9\,K sections, the much larger interaction cross-sections of the other gas species 
causes them to dominate the interaction rate, especially in the LSS. Nitrogen is therefore considered to represent a 
good average of the atomic composition of the residual gas\,\cite{nima729:825}.
Using a generic gas species and a uniform distribution of events has the advantage 
that the same simulation results can be used with different pressure distributions.
Unlike most previous studies\,\cite{nima381:531,nikolai1, nikolai2, lhcb-ieee}, all simulations 
in this work were performed without any Monte Carlo variance reduction techniques,\footnote{Variance reduction
refers to favouring some regions of phase space at the cost of others in order to achieve faster convergence
of the estimates in the favoured regions. While significantly reducing the computational effort, the disadvantage of 
these methods is that they do not preserve correlations within events.} in order to preserve correlations within 
individual events. This is a prerequisite for reconstructing the trigger signatures.
In the first simulation step, the secondaries produced in the beam--gas interactions are transported to a 
virtual {\em interface plane} at $z\!=\,$22.6\,m upstream of the IP. The choice of this $z$-location is motivated 
by the fact that it is on the IP-side of the closest inner-triplet magnet. Thus it naturally separates the 
experimental area, where a detailed \fluka geometry of the ATLAS detector is available, from the 
LHC accelerator with its own geometry and magnetic field modelling.
For all particles reaching this plane the positions, four-momenta and times of flight are recorded 
and serve as input to subsequent detector simulations. 
For a complete description of the background, especially in the BCM with its low hit threshold, 
particles have to be transported down to low energies to ensure that all
potential hits are simulated. Since this is very CPU-intensive, only six million inelastic events 
were simulated, transporting particles down to a kinetic energy of 20\,\MeV.
This value of 20\,\MeV\ is chosen in order to stay above low-energy nuclear reactions
which are the source of a large number of low-energy particles. Such particles are 
absorbed locally but, due to their abundance, their simulation is costly in terms of 
CPU time. Six million events are not enough for the fake-jet studies, so a second 
sample of 300 million $p$--N$_2$ interactions was generated with a threshold of 20\,\GeV.\footnote{In order to
further increase the number of fake jets, all events were used twice. Since the probability for a muon to experience 
a large radiative energy loss is very low, only 5\% of the events gave a fake jet on both trials. Even in these 
cases the jet \pt was different, i.e.\ only the azimuthal angle was strongly correlated.}
This 20\,\GeV\ threshold corresponds to the minimum energy of the muon needed to create a fake jet with 
sufficient transverse momentum to fire the jet trigger used in this study.

\begin{figure}[t]
\centering
\includegraphics[width=0.98\textwidth]{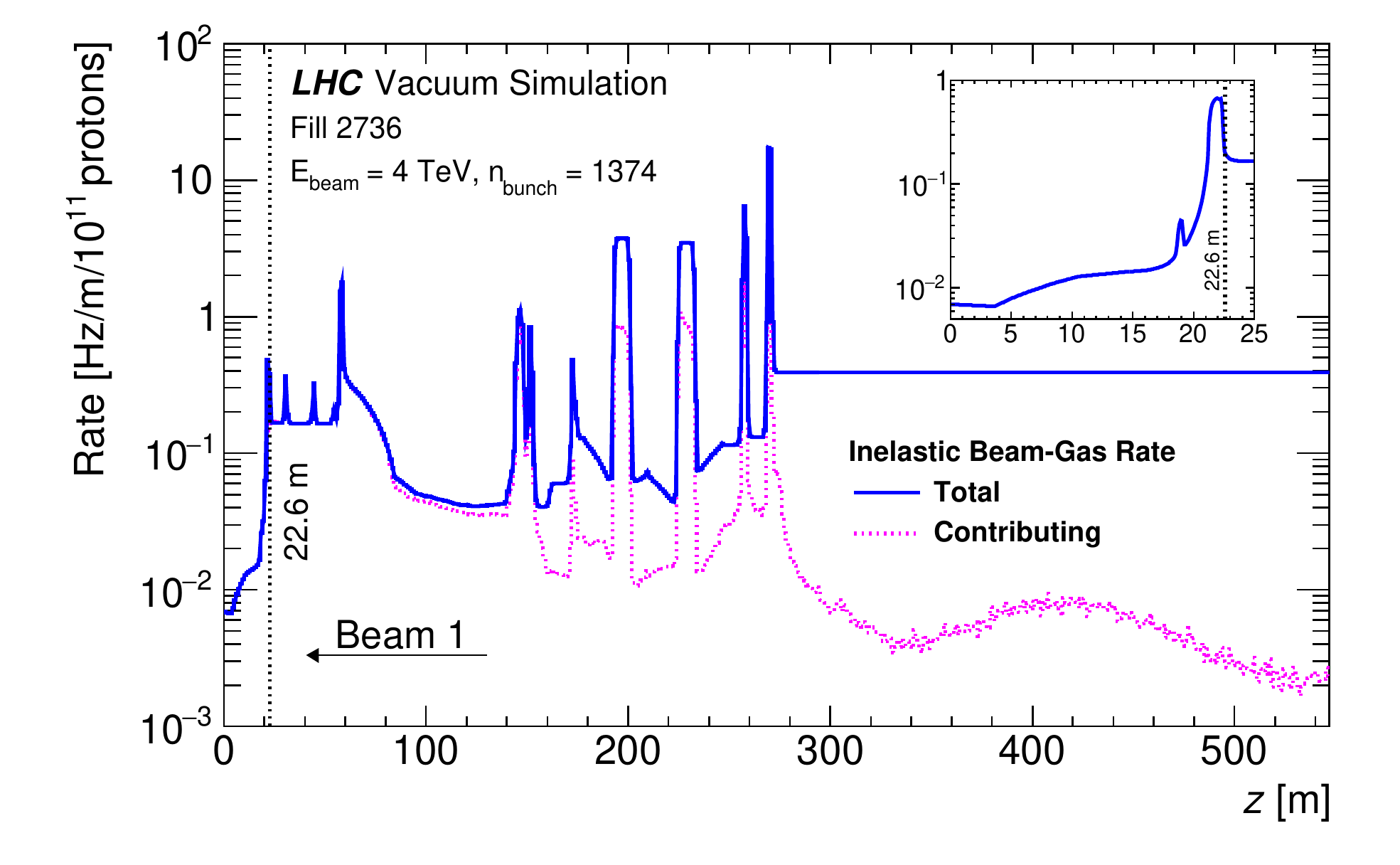}
\caption{Inelastic beam--gas interaction rate of beam-1 in IR1 as a function of distance from the IP at the
start of data-taking in LHC fill 2736. The beam moves towards negative $z$, i.e.\ from right to left in the figure.
The total rate (solid blue histogram) reflects the residual gas pressure. The dotted histogram
shows the rate of interactions which contribute at least one particle with kinetic energy $E>20$\,\MeV\ at the interface 
plane at $z = 22.6$\,m.  
The prominent peaks between $z\approx 150$\,m and $z\approx 270$\,m correspond to the positions of the TCT, the D2 dipole (T = 4.5\,K), 
Q4--Q6 quadrupoles (T = 4.5\,K) and cold-warm transitions at the exit of the arc. 
The pressure in the LHC cold arc (T = 1.9\,K), starting at $\sim 270$\,m, is assumed constant. The small inset shows the 
interaction rate on the IP side of the interface plane in more detail.}
\label{fig:pint}
\end{figure}

The rate of $p$--N$_2$ interactions as a function of $z$, shown by the solid histogram in Figure\,\ref{fig:pint},
is obtained from an equivalent N$_2$ density distribution of the residual gas, $\rho_{\mathrm{N}_2}(z)$. The 
partial densities ($\rho_i$) of all residual gas species at the location $z$ are taken from
the simulated pressure distribution and 
weighted by the ratio of inelastic proton--molecule ($\sigma_i$) to $p$--N$_2$ ($\sigma_{\mathrm{N}_2}$) 
cross-sections:
\begin{equation}
\rho_{{\mathrm N}_2}(z) = \sum_i\frac{\rho_i(z)\cdot\sigma_i}{\sigma_{\mathrm{N}_2}},
\label{eq:xsecnorm}
\end{equation}
where $i$ runs over H$_2$, CH$_4$, CO and CO$_2$.
The absolute normalisation of all simulated rates in this paper is fixed by the interaction rates
shown in Figure\,\ref{fig:pint}. Since these are derived from the pressure 
distribution, they are subjected to the uncertainty in the pressure simulations, discussed in Section\,\ref{sect:bib}.

The events used as input to the ATLAS simulations are sampled according to their $z$-coordinate, using the rate 
distribution of inelastic interactions, shown in Figure\,\ref{fig:pint}. 
The dotted histogram in Figure\,\ref{fig:pint} shows, as a function of $z$,
the rate of those events for which at least one particle
has reached the interface plane. A comparison of the two histograms 
in Figure\,\ref{fig:pint} reveals that practically all events produced at $z\lesssim 150$\,m give 
contributions, while only $\sim$1\%  of events with $z\!>\!300$\,m result 
in particles at the interface plane.  

In order to account for beam--gas events between the IP and the interface plane, 
$p$--N$_2$ events were generated separately for $z\!<\,$22.6\,m 
with a $z$-distribution sampled directly from the inelastic interaction probability
in that region, i.e.\ left of the dashed vertical line in Figure\,\ref{fig:pint}.

The particles at the interface plane, as well as those generated 
at $z\!<\,$22.6\,m, were transported through the ATLAS experimental area and detector using a 
dedicated \fluka geometry model\,\cite{Baranov:2005ewa}. The magnetic fields produced by the ATLAS 
magnets were implemented as two-dimensional maps covering the entire detector radius and extending
in $z$ up to the interface plane.
The propagation and showering of the particles through ATLAS was simulated with \fluka,
which provides accurate simulation of all relevant physics processes. Besides 
full simulation of hadronic and electromagnetic showers, \fluka provides detailed
transport of muons through matter with complete modelling of all energy loss processes and
explicit production of secondary particles in radiative events.
Compared to the full ATLAS simulation\,\cite{atlassoft} based on \GEANT4\,\cite{geant4},
the disadvantage of choosing \fluka is that an exact modelling of the detector response is
not available. In particular, digitisation and reconstruction of e.g.\ tracks and jets cannot be 
performed in \fluka simulations with a level of detail comparable to real data.
Dedicated algorithms were incorporated in the \fluka simulation in order to record 
quantities of interest, namely energy depositions and detector hits, on an event basis. The 
rates of fake jets and events with the  BCM background trigger signature were estimated using 
custom reconstruction algorithms during the post-processing of the simulation output.

The geometry of the BCM detector was modelled, including both the sensitive detector and the services.
The transport threshold in the ATLAS simulations was set to 100\,\keV, 
so that all particles able to generate hits in the BCM detector 
were included in the simulations. Neutrons were always transported to thermal energies
and their capture by nuclei, with associated photon emission, was simulated.

Due to the 20\,\MeV\ transport threshold the LHC simulations do not include particles down to 100\,\keV. 
This has no significant influence on the results, since particles starting from the interface plane 
will not reach the BCM directly: most of them are intercepted by the TAS. Those which pass through its 
small central aperture have to traverse the beam-pipe wall at a very shallow angle, which implies a 
high probability for an inelastic interaction. 
This was verified by checking that the BCM trigger rates as a function of the $z$-coordinate of the origin of the 
event, as obtained from the ``fully 100\,\keV'' simulation of events on the IP side of the interface plane, 
join smoothly with the rates from the ``mixed 20\,\MeV\ \& 100\,\keV'' simulation of events beyond $z = 22.6$\,m.
In the simulations the threshold of a BCM module is accounted for by considering as a hit each charged particle with a kinetic energy 
above 250\,\keV, entering the sensitive area of a BCM module. This simplification is motivated by the fact that a particle 
deposits in the module at least the minimum-ionising equivalent or its total kinetic energy, whichever is smaller. 
The sensitivity of the results to the choice of threshold was evaluated by varying it between 100\,\keV\ and 1\,\MeV. The 
simulated BCM trigger rate was affected by only a few percent. The arrival time and the identifier of the 
module entered are used to reconstruct the BCM background triggers from the recorded data and simulation output.
Each hit results in a dead time of the affected BCM module, the duration of which depends on the
energy but is typically 10--20\,ns. For simplicity only the first hit in each BCM module, in a $\pm12.5$\,ns window 
around $t=0$, is considered, both in the simulations and the data. 

\begin{table}[t]
\renewcommand{\arraystretch}{1.3}
\caption{Radial and longitudinal extent of the ATLAS calorimeter regions and bin sizes ($\delta r$, $\delta z$) as implemented in 
the \fluka geometry. 
An azimuthal binning of 36 bins of 10 degrees each is used in all calorimeter regions.
The endcaps at the negative (`$-$') side of ATLAS are mirror images of the positive (`$+$') ones.
}
\label{tabCalo}
\centering
\begin{tabular}{l|c|c|c|c|c|c}
\hline
            & $r_{\textrm{min}}$ & $r_{\textrm{max}}$ & $z_{\textrm{min}}$       & $z_{\textrm{max}}$ & $\delta r$  & $\delta z$  \\
Calorimeter & [mm]      & [mm]      & [mm]            & [mm]      & [mm]  & [mm]         \\ \hline
Barrel LAr  & 1471      & 2009      & $-$3172           & 3172      & 107.6 & 396.5 \\
Barrel Tile & 2285      & 3885      & $-$6000           & 6000      & 160.0 & 400.0 \\
Endcap1 ($+$) & \phantom{0}475       & 2075      & \phantom{0}3670 & 6120      & 160.0 & 408.3 \\
Endcap2 ($+$) & \phantom{0}300       & \phantom{0}475       & \phantom{0}3670 & 4650      & \phantom{0}87.5  & 490.0 \\ \hline
\end{tabular}
\end{table}

\begin{figure}[t]
\centering
\includegraphics[width=1.05\textwidth , trim=2.0cm 0cm 0cm 0cm, clip=true]{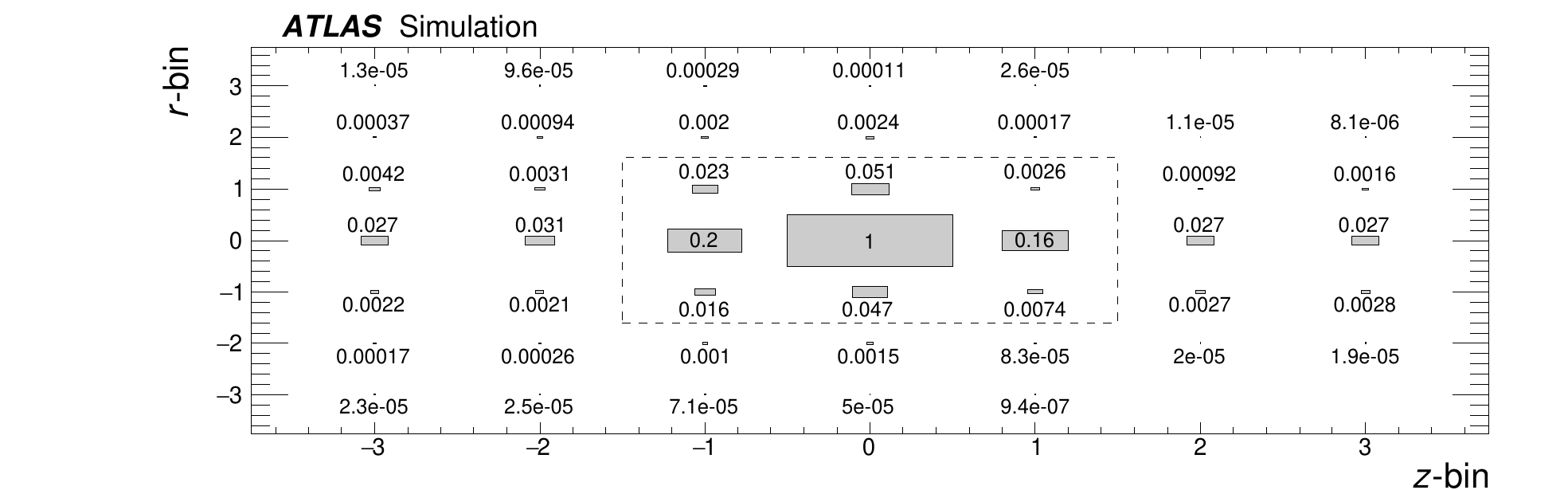}
\caption{
The $r$--$z$ projection of the average fractional energy distribution, found in the \fluka simulations 
for jets with \pt$>$16\,\GeV\ in the barrel LAr calorimeter. The beam direction is from right 
to left in the plot. The values are summed over seven bins in azimuth, centred at the maximum, and 
normalised such that for each event the maximum bin is 1.0. Before averaging over all 
events, they are aligned such that the largest deposition is at the centre of the plot. The $r$ and $z$
bin numbers shown in the plot are relative to the centre. The dashed box indicates the $3\!\times\!3$ $r$--$z$ 
bins used for determining the jet energy (summed over azimuthal bins).} 
\label{edr_clusters}
\end{figure}

Since fake jets are mainly produced by radiative energy losses of high-energy muons, the
computational effort was significantly reduced by selecting only muons from the sample with a 20\,\GeV\ threshold.
The propagation and showering in ATLAS was simulated with a 100\,\keV\ transport threshold.
The fake-jet rates are estimated by recording, event by event, the local energy depositions 
in the different calorimeter regions which are described in Table\,\ref{tabCalo}. 
After the simulation of each event, the energy depositions are analysed. 

Since \fluka is not part of the standard ATLAS simulation software, it does not
benefit from the sophisticated ATLAS jet-reconstruction tools. Instead, a much simplified
algorithm is used to assess the fake-jet rate in the simulations.
A cluster is formed by summing the energy depositions in $3\!\times\!3\!\times\!3 = 27$ ($r$, $\phi$, $z$)-bins, 
centred around the maximum deposition. In the barrel calorimeters this clustering produces
jets with angular dimensions comparable to those of the ATLAS jet-reconstruction algorithm.
Each deposition is used only once, starting with the highest in any bin.
If the cluster energy is large enough to exceed the 10\,\GeV\ transverse-energy threshold, 
the cluster is counted as a fake jet. The position of the fake jet is determined 
from the energy-weighted average of the bins considered. Likewise the jet time is determined 
from the energy-weighted time of the individual depositions. 
Depositions at times larger than 50\,ns are excluded
in order to prevent small depositions with very large delay, e.g.\ from thermal neutron capture, 
to influence the average time. This procedure is fully consistent with the reconstruction of jet time in ATLAS data, 
which also takes into account only depositions in a narrow time window.

In order to assess the systematic uncertainty due to the energy spread, sums over more bins were explored and
it was found that an extension of the sum in $r$ and $\phi$ adds almost no energy to the cluster.
In $z$, however, taking the sum over more bins results in a larger cluster energy. 
Figure\,\ref{edr_clusters} shows the $r$--$z$ projection of the average energy fraction in the different bins 
around the maximum. 
The energy is well contained in the central $3\!\times\!3$ bins. 
The continuous energy loss of the passing muon, about 1\,\GeV\ per metre in the calorimeters, is reflected as a 
row of almost constant values for $r$-bin\,=\,0 and $|z|{\textrm {-bin}} > 1$ in Figure\,\ref{edr_clusters}. 
A wider summing range in $z$ mostly adds this ionisation energy loss of the muon, which would be a non-negligible 
contribution to the lowest jet energies considered in this study.
The average energy lost by the muon, however, is below the threshold of the ATLAS jet reconstruction, so in data only 
upward fluctuations of the muon energy loss are likely to be combined into the jet, if they happen close enough to the large 
radiative loss. Therefore an energy sum over $3\!\times\!3\!\times\!3$ bins is considered a good approximation to the 
reconstruction algorithm applied to the data.

\section{Comparison with data}

The principal objective of this work is to validate, through comparisons with data, the simulation methods described in Section\,\ref{sect:simulations}. 
For this purpose, events collected with the BCM background and low-\pt jet triggers during 2012
are analysed. The vacuum simulations assume the beam conditions at the start of LHC fill 2736, which correspond to
the parameters listed in Table\,\ref{tabLHC} and are typical of the operation in the second half of 2012. 
Only fills with the same bunch pattern as in fill 2736 are considered in the analysis. Data affected by more 
than 20\% trigger dead time are rejected and a dead-time correction is applied to the remaining data.
In order to remove the effect of the beam intensity, which decreases in 
the course of a fill, all results are normalised to $10^{11}$\,protons. However, since the residual gas 
pressure follows the decrease of beam intensity over a LHC fill, fill-averaged beam--gas rates are lower 
than those at the start of a fill. 

\subsection{BCM background}

\begin{figure}[t]
\centering
\includegraphics[width=0.9\textwidth]{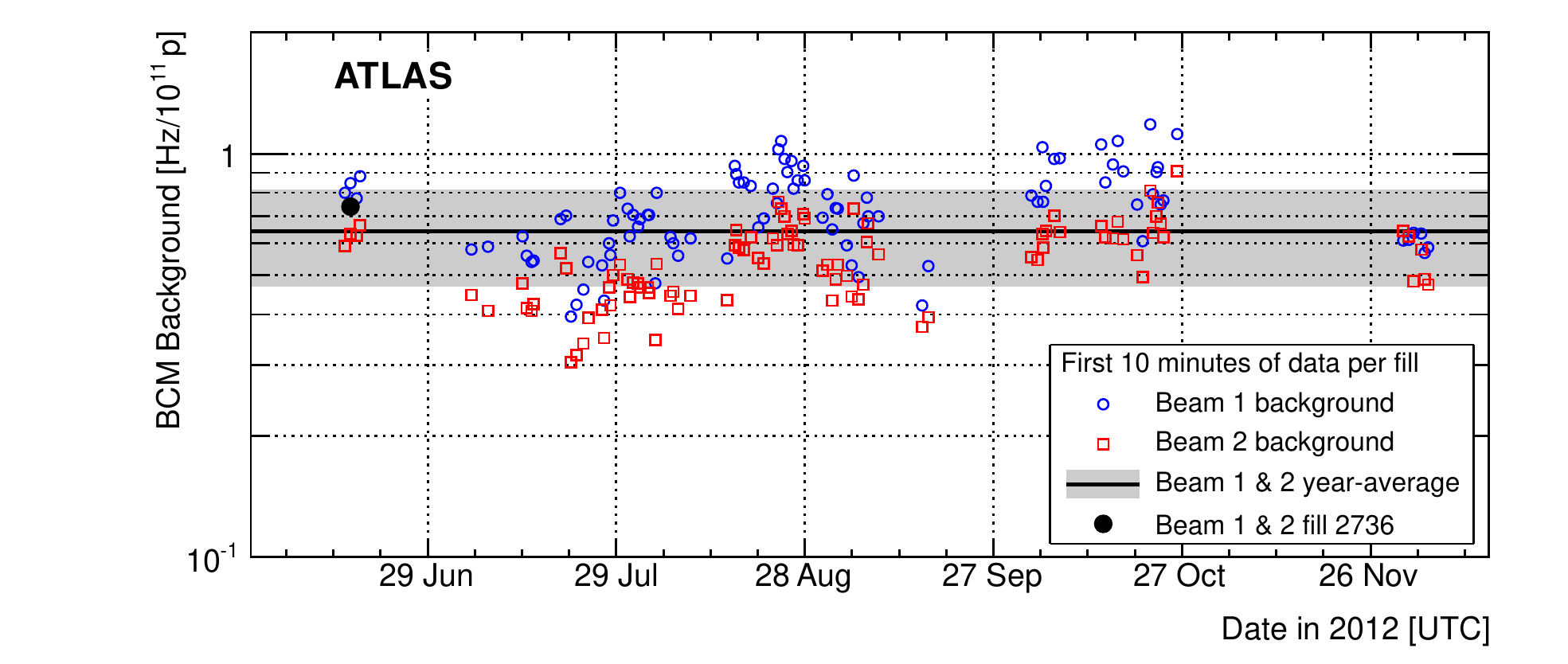}
\caption{BCM background rate during the first ten minutes of data-taking in each fill with 1374 bunches.
The average and standard deviation, resulting from the combined effect of fill-to-fill variation and difference between the beams, are 
shown by the black line and shaded area respectively. Data taken during a period with instrumental noise in the BCM, lasting through most of 
November, are excluded\,\protect\cite{backgroundpaper2012}.
Only fills where data-taking started promptly after beams were brought into collision and which provided at least ten minutes of data are included. 
The solid black circle in mid June shows the average of beam-1 and beam-2 backgrounds in fill 2736, for which the pressure distribution 
has been simulated.
}
\label{fig:bcmOverYear}
\end{figure}

In Figure\,\ref{fig:bcmOverYear} the BCM background rates during the first ten minutes of data-taking
are shown for all LHC fills included in the analysis. The direction information provided by the BCM is used 
to reject events in the direction opposite to the unpaired bunches, which are used for the background measurement. 
Such wrong-direction signals  can arise either from {\em ghost charge}\,\footnote{Ghost charge is formed 
by beam protons that have escaped their initial RF-bucket and been recaptured in nominally empty buckets.} 
in the opposite beam or from accidental background signatures involving hits from {\em afterglow}\,\cite{backgroundpaper2012}.
Although the data are selected such that they should correspond to the same beam conditions, a significant fill-to-fill
variation and slightly increasing trend over the year can be seen. 
The BCM background from beam-1 is found to be systematically higher than from beam-2. The relative difference,
averaged over all the data in Figure\,\ref{fig:bcmOverYear}, is 28\%. Since the simulations make no 
distinction between the two beams, they are compared with the average. Although, at $\pm$14\%, the difference between the 
beams is small compared to the fill-to-fill variation, it is included in the variation in the data quoted in Table\,\ref{tabRateBCM}. 

Table\,\ref{tabRateBCM} compares the simulated BCM beam background rates with the start-of-fill 
and the fill-averaged data taken in 2012.  
The simulated rate of 1.2\,Hz/10$^{11}$\,protons is almost twice the measured start-of-fill value.
Figure\,\ref{fig:bcmOverYear} shows separately the observed BCM background rate in fill 2736, for which the 
pressure simulations are performed. With a rate of 0.72\,Hz/10$^{11}$\,protons it falls close to the upper edge of the fill-to-fill 
variation and thus closer to the simulated value than the 2012 average shown in Table\,\ref{tabRateBCM}.

In the inner triplet, $p$--H$_2$ scattering contributes about 90\% of the beam--gas interactions, 
while the simulations are based on $p$--N$_2$ events. Equation\,(\ref{eq:xsecnorm}) ensures that the
correct number of beam-gas collisions is generated in the simulations, but it does not account for 
differences in the collision dynamics, especially the multiplicity of produced secondaries.
In order to estimate the possible
dependence of the background rate on the target nuclide, the less CPU-intensive simulations at $z<22.6$\,m  
were repeated with $p$--H$_2$ events. The rate was found to decrease by about 15\%. Assuming a similar reduction for
the inner-triplet region, where most of the BCM background originates from (see Section\,\ref{sect:zorigin}), the use of 
proton--N$_2$ events overestimates the BCM trigger rate by up to 15\%.

Even after accounting for this correction, the observed difference between simulation and data is larger than the fill-to-fill variation, but remains well within 
the estimated uncertainty range of the simulation, which is dominated by knowledge of the pressure distribution.

\begin{table}[t]
\renewcommand{\arraystretch}{1.3}
\caption{
Simulated BCM background rates compared with ATLAS data. The rates correspond to events giving, in the BCM, 
a background signature that is consistent with the direction of the unpaired bunch. The simulations correspond to the 
start of data-taking, while the last two columns illustrate the difference in background between averaging over the first 
ten minutes of data-taking in each fill and averaging over entire fills.
For the data, the uncertainty in the average corresponds to one standard deviation of the mean of all fills. 
For the simulations it indicates the statistical uncertainty.
The fill-to-fill variation includes the difference between beam-1 and beam-2. 
The last row indicates the possible range of the simulated rate, due to the estimated uncertainty 
of the pressure simulation, discussed in Section\,\protect\ref{sect:bib}.
}
\label{tabRateBCM}
\centering
\begin{tabular}{l|c|c|c} \hline
                                              &  MC simulation         &  Data                            & Data                             \\ 
                                              & [0--546.6]\,m          &  Fill Start                      & All Fill                         \\ \hline
Average rate [Hz/$10^{11}$\,protons]          & 1.2                    &  0.642                           & 0.463                            \\ 
Uncertainty in average rate                   & 0.4\%                  &  2.0\%                           & 2.3\%                          \\
Fill-to-fill rate variation                   & ---                    &  27\%                            & 34\%                           \\ \hline
Pressure uncertainty [Hz/$10^{11}$\,protons]  & 0.4--3.6               &  ---                             & ---                              \\ \hline

\end{tabular}
\end{table}

\begin{figure}[t]
\centering
\includegraphics[width=0.7\textwidth, trim=0cm 3.5cm 0cm 0cm, clip=true]{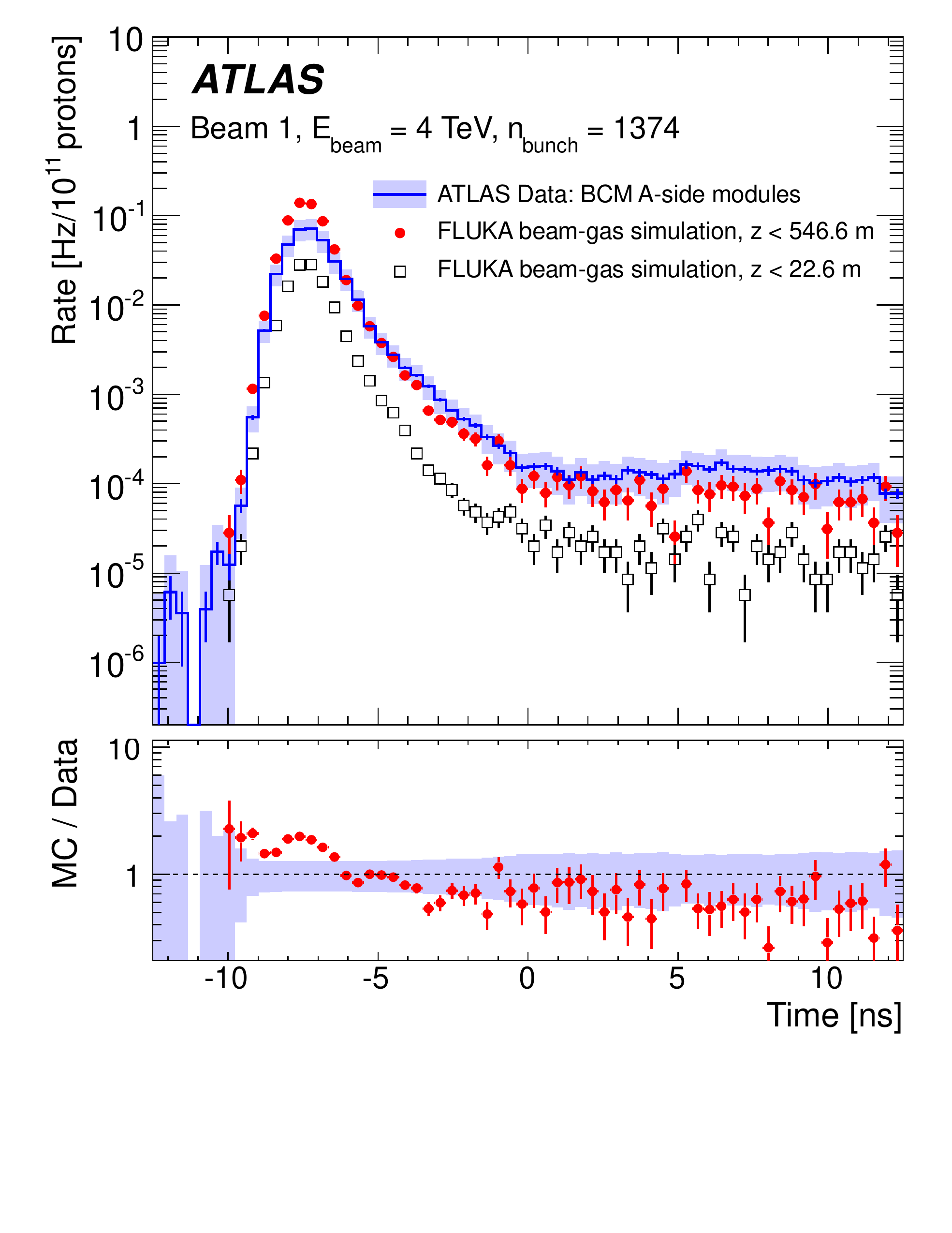}
\caption{
Comparison of the time distribution of BCM hits in ATLAS unpaired-bunch data (blue histogram) with \fluka simulation (red circles). 
The bunch passes the IP at $t\!=\,$0\,ns.
The histograms show the early hits per upstream BCM module for events which have fired the BCM beam-background trigger 
in the beam-1 direction. 
The black open squares show the contribution from beam--gas events within the ATLAS experimental area $z<22.6$\,m,
while the red solid circles show the total, i.e.\ $z<546.6$\,m. The errors shown on the simulation are statistical only.
The blue band indicates the fill-to-fill and module-to-module variation of the data and the error bars show the
uncertainty in the mean value. The lower panel shows the ratio of simulation to data, taken between the red circles 
and the blue histogram.
Only data from the first ten minutes of ATLAS data-taking in each LHC fill are considered.
}
\label{fig:bcmTime}
\end{figure}

In Figure\,\ref{fig:bcmTime} the distribution of particle arrival times at the BCM modules in 
the simulation is compared with data. 
The histograms represent the time distribution of hits in upstream BCM modules  
for events which give the BCM background signature in beam-1 unpaired bunches. 
The plain \fluka simulations yield a very narrow time distribution with a vertical rising edge. In the analysis, this is smeared by 
the 0.25\,ns time fluctuation due to the LHC bunch length of 75\,mm.\footnote{The centre of the bunch passes the IP at $t$ = 0, but the background event can
originate from any proton within the bunch.} A larger broadening effect comes from the instrumental resolution
and time alignment of the BCM. In order to account for these, 
the rising edge of the simulated time distribution is fitted to that in data with the time alignment and time
resolution as free parameters. Values of --1.0\,ns and 0.55\,ns are found for these parameters respectively.  
The fit yields an uncertainty of about 10\% in both parameters. The observed time shift is well within the 
2\,ns alignment tolerance specified for the BCM. 
The fitted time resolution is about 30\% better than that found in test-beams\,\cite{bcm}. 
However, it agrees, within its 10\% uncertainty, with the resolution derived from in-situ monitoring of
the time difference between hits in upstream and downstream modules in collision events recorded by the BCM detector.

The data shown in Figure\,\ref{fig:bcmTime} are extracted from the events recorded during the first ten minutes of each 
LHC fill.
A determination of the fill-to-fill variation for each bin is not feasible since, especially in the tail region, the 
very low counting rate causes many bins to have zero counts for a single fill. If, however, the shape of the distribution 
is assumed to be invariant between fills, the fill-to-fill variation of each bin can be taken as 27\%, which is the value
given in Table\,\ref{tabRateBCM} for the total rate. The blue band shown around the data 
in Figure\,\ref{fig:bcmTime} illustrates this fill-to-fill variation, but takes also into account the 
total counting statistics in each bin. The uncertainty in the mean value in each bin is determined from the data in 
all fills and shown by the smaller error bars on the data. For the simulations, only the statistical uncertainties are shown.
The error bars on the ratio of simulation to data are based on the statistical uncertainties only.
If the shape of the distribution is correctly reproduced by the simulation, the ratio should be a constant. However, since the 
simulations correspond to a particular fill, this constant can deviate from unity by the amount of the fill-to-fill variation. The 
allowed range ($1\sigma$), ignoring the uncertainty arising from the pressure distribution, is indicated by the blue band.

The ratio shown in Figure\,\ref{fig:bcmTime} indicates that the peak is overestimated by the simulations, while the 
tail at positive times is underestimated, i.e.\ the peak-to-tail ratio is larger in the simulations than in data and, 
consequently, the falling slope is slightly steeper. 
The open squares in Figure\,\ref{fig:bcmTime} show that the events within the ATLAS experimental area contribute
only 20\% to the total hit rate but the shape, especially the peak-to-tail ratio, is similar to that of the total rate.
If the tail were due to delayed arrival of some particles from distant events, i.e.\ due to a dependence between time
spread and distance to the event, then it should not appear for the beam--gas events at $z<22.6$\,m.
The fact that a tail of similar height, relative to the peak, is seen in both distributions indicates that the delayed 
particles are due to a local effect. The simulated hits in the tail are found to be caused by particles with a kinetic 
energy below $\sim$10\,\MeV\ and if the solenoidal field is turned off in the simulations, the tail is suppressed. 
These findings indicate that the tail is due to low-energy particles looping in the magnetic field and
accumulating a delay due to the longer path along the helix. Since such low-energy particles have a short range in matter
already a slightly too large amount of material in the simulation geometry, with respect to reality, is sufficient to 
explain the observed underestimation.

\begin{figure}[t]
\centering
\includegraphics[width=0.8\textwidth]{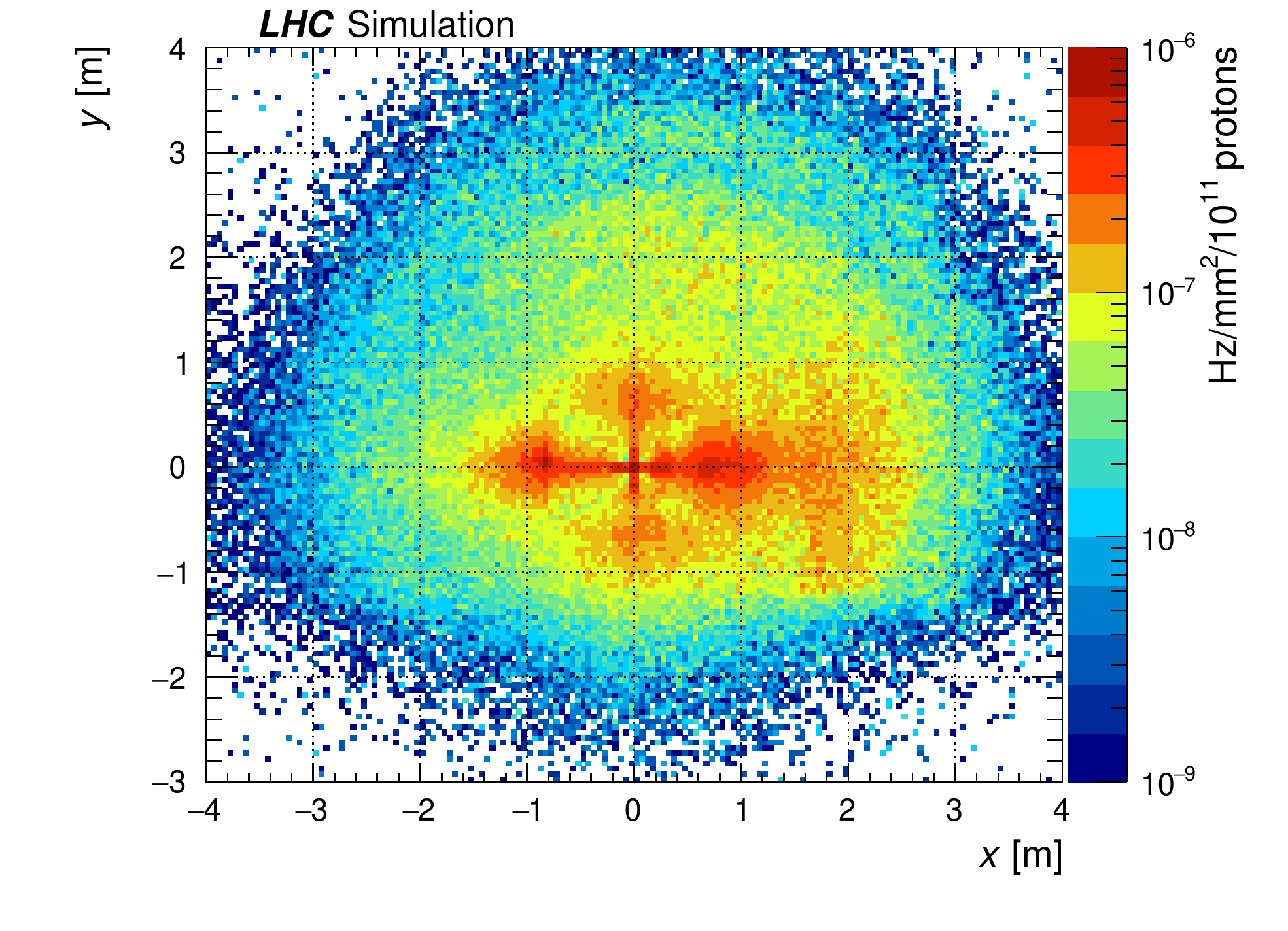}
\caption{
Simulated $x$--$y$ distribution of muons with energy $E>$20\,\GeV\ entering the ATLAS experimental area at $z\!=\,$22.6\,m. 
The beam passes at $(x,y)\!=\!(0,0)$. The plot is based on Ref.\,\protect\cite{regina} and constitutes the
input to the study described in this paper. The rate corresponds to the pressure conditions at the start of fill 2736.
}
\label{fig:muonsIntPlane}
\end{figure}

\subsection{Fake-jet background}

Most of the fake jets are produced by radiative energy losses of high-energy muons in the 
calorimeters. Such fake jets have a very different topology from collision jets: 
they do not point to the IP and are almost entirely of electromagnetic nature with very little hadronic activity. 
Therefore the simulation results are compared with jet data calibrated to the electromagnetic energy scale 
rather than with fully calibrated jets, which are corrected for the non-compensating response of the calorimeter to hadrons. 
Possible jets from collisions of the protons in the unpaired bunch with ghost charge 
in the other beam are removed by rejecting events for which a primary vertex has been reconstructed from the
tracks measured by the inner detector. Only the highest-\pt jet in each event is included in the analysis.
In the endcap calorimeters, hadronic showers can contribute to the fake jets. Since only muons are considered in the
simulations, the analysis is restricted to $|\eta|<1.5$, i.e.\ the barrel calorimeters. These are at
large radii and shadowed by other detector elements so hadronic showers from the beam line cannot reach them.

Figure\,\ref{fig:muonsIntPlane} shows the $x$--$y$ distribution of  muons with energy $E>20$\,\GeV\ reaching the interface plane.
This distribution reflects the geometry of the LHC tunnel, and also the effect of some beam-line magnets.  
The tunnel's radius of 2.2\,m and the floor at $y\!=\,$--1.1\,m produce a relatively sharp edge in the 
muon flux. The higher rate seen on the inside of the ring at \mbox{$y\!\approx\!\pm$1\,m}, between $x\!\approx\,$1.5\,m and $x\!\approx\,$2.5\,m, 
is due to the offset of the beam line relative to the centre of the tunnel, leaving  more free space for pions and kaons to decay 
into muons on the inside of the ring.
The ``hot spots'' seen around $x\!\approx\!\pm$0.8\,m are mainly due to bending of the off-momentum muons by the D1 and D2 dipoles 
of the LSS. The vertical spread at $x=0$ is probably due to bending in the quadrupoles of the inner triplet, although the crossing 
angle might also have some influence on this.

In Table\,\ref{tabRateFJ} the rates of simulated fake jets, created by the muons shown in Figure\,\ref{fig:muonsIntPlane}, are compared  
with the data from all relevant fills in 2012.
Systematic uncertainties may arise from the jet reconstruction used in the simulations.
The studies described in Section\,\ref{sect:simulations} show that increasing the extent over which the jet energy is 
integrated in the simulations from $3\!\times\!3\!\times\!3$ bins to a very wide 
$7\!\times\!7\!\times\!7$ bins increases the jet rate by 20\%. However, since this increase is due to 
including the ionisation energy loss of the passing muon, it does not seem justified to consider it as a systematic 
uncertainty, but rather an upper limit thereof. Thus the uncertainty from the jet reconstruction algorithm is 
considered negligible compared with the uncertainty from the pressure distribution.
The latter is estimated to be a factor of three, which means that the high level of agreement between simulations and
data, seen in Table\,\ref{tabRateFJ}, must be largely fortuitous.

\begin{table}[t]
\renewcommand{\arraystretch}{1.3}
\caption{
Simulated fake-jet rates compared with ATLAS data. Only jets with \pt$>$16\,\GeV\ and $|\eta|<1.5$ are considered.
The simulations correspond to the start of data-taking, while the last two columns 
illustrate the difference in background between
averaging over the first ten minutes of data-taking in each fill and averaging over entire fills.
For the data, the uncertainty in the average corresponds to one standard deviation of the mean of all fills. 
For the simulations it indicates the statistical uncertainty.
The fill-to-fill variation includes the difference between beam-1 and beam-2. 
The last row indicates the possible range of the simulated rate, due to the estimated uncertainty 
of the pressure simulation, discussed in Section\,\protect\ref{sect:bib}.
}
\label{tabRateFJ}
\centering
\begin{tabular}{l|c|c|c} \hline
                                               &  MC simulation         &  Data                            & Data           \\ 
                                               & [0--546.6]\,m          &  Fill Start                      & All Fill       \\ \hline
Average rate [Hz/$10^{11}$\,protons]           & 0.0053                 &  0.0046                          & 0.0037         \\ 
Uncertainty in average rate                    & 1.0\%                  &  3.4\%                           & 2.2\%          \\
Fill-to-fill rate variation                    & ---                    &  56\%                            & 39\%           \\ \hline
Pressure uncertainty [Hz/$10^{11}$\,protons]   & 0.002--0.015           &  ---                             & ---            \\ \hline
\end{tabular}
\end{table}

\begin{figure}[t]
\centering
\includegraphics[width=0.5\textwidth]{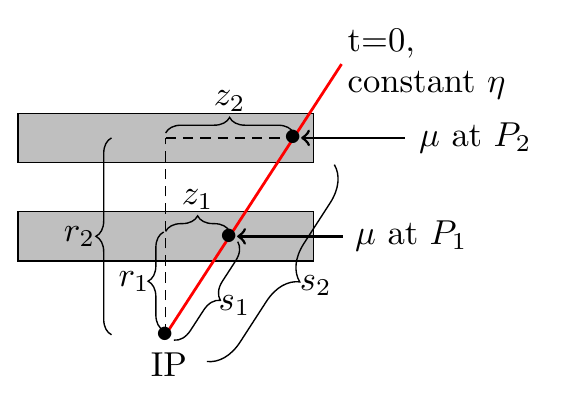}
\caption{
Schematic illustration of reconstructed arrival time of beam background muons 
compared with collision jets in the calorimeter regions.
The two grey boxes represent two different calorimeters  
and the red solid line represents $t\!=\,$\,0, corrected for 
the time of flight of particles coming from the IP.
}
\label{fig:schematicFakeJetTime}
\end{figure}

\begin{figure}[t]
\centering
\subfloat[]{\includegraphics[width=0.48\textwidth]{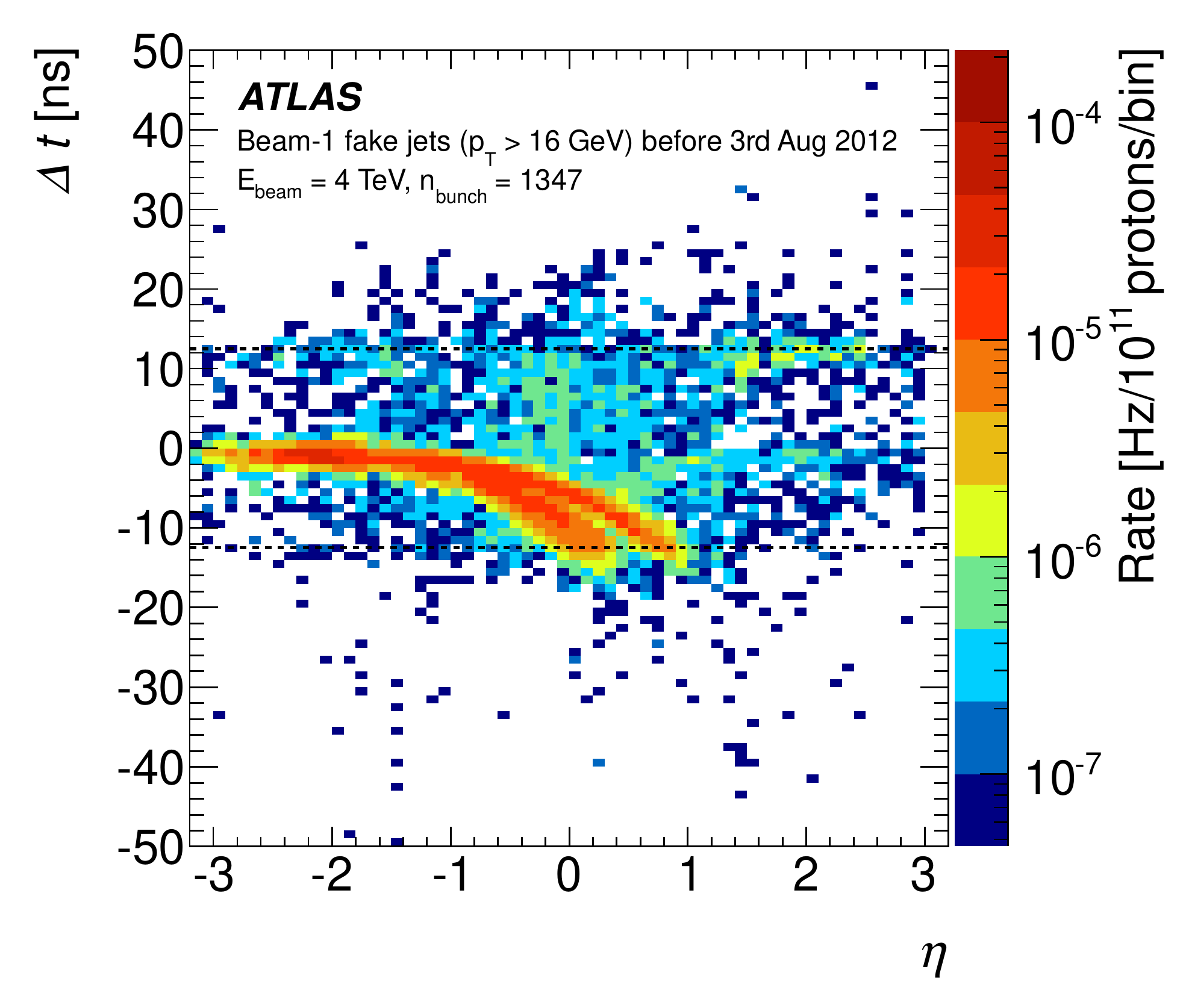}
\label{fig:bananaDat}}
\subfloat[]{\includegraphics[width=0.48\textwidth]{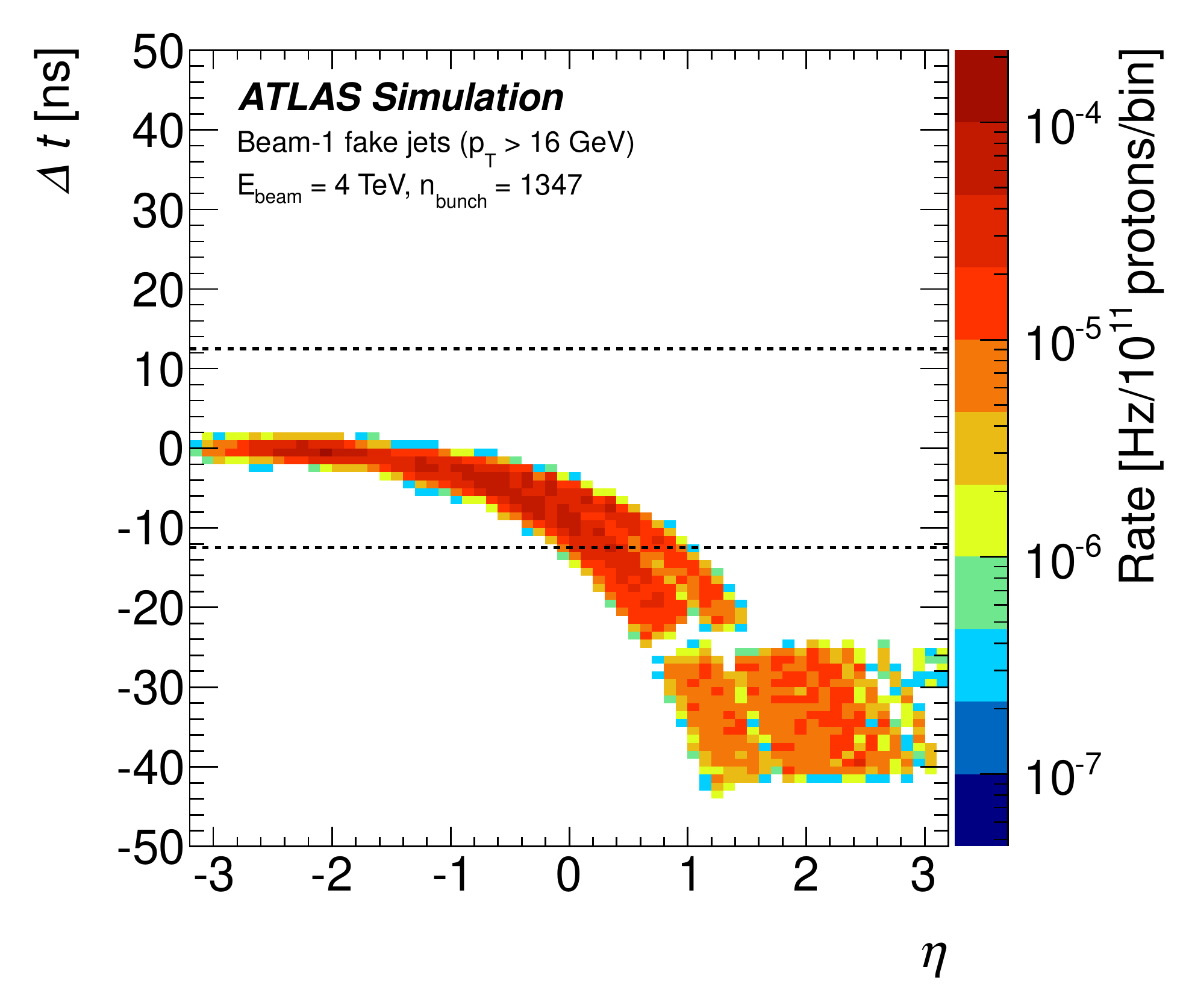}
\label{fig:bananaSim}}
\caption{
(a) Fake-jet counts in unpaired bunches in the pseudorapidity--time ($\eta$--$\Delta t$) plane for beam-1 in ATLAS data
and (b) the simulated rate of energy deposition clusters with \pt\!$>$16\,\GeV\ in the $\eta$--$\Delta t$ plane. 
The width of the bins is 1\,ns in time and 0.1 units in $\eta$. 
The \fluka simulations, which correspond to the start-of-fill conditions, have been scaled by the ratio of the 
total "All Fill" to "Fill Start" rates, shown in Table\,\ref{tabRateBCM}.
}
\label{fig:bananas}
\end{figure}

The offset in arrival time, at given $z$, between the proton bunch and a beam background muon originating from that bunch, 
is negligible. Therefore, when the beam background muon reaches an upstream point $P\,(r, z)$ in the calorimeter, 
the proton bunch still has to cover a distance $|z|$ to reach the IP and then the produced secondary 
particles have to travel a distance $s = \sqrt{r^2+z^2}$  to reach $P$, 
as illustrated in Figure\,\ref{fig:schematicFakeJetTime}. 
For a downstream $P$ the expression for $s$ is the same,
but in this case the muon has to cover the additional distance $|z|$.
The calorimeter timing is such that for each point $P$ the arrival time of a secondary particle produced 
in collisions at the IP is 0. 
Thus the relative time $\Delta t$ of a beam background muon at $P$ is given by  
\begin{equation}
\Delta t = - \left(\sqrt{r^2+z^2} \pm |z| \right)\!/ c
\label{eq_jetTime}
\end{equation}
where $c$ is the speed of light 
and $+|z|$ and $-|z|$ correspond to the upstream and downstream sides respectively. 
Equation\,(\ref{eq_jetTime}) shows that fake jets due to BIB always arrive early, i.e.\ have $\Delta t < 0$. 
A characteristic {\em banana} shape is seen in the $\eta$--$\Delta t$ plane, shown in Figure\,\ref{fig:bananas}; 
this arises from the definition of $\eta$ and the dependence of jet time on $z$ and $r$.

The number of fake-jet counts in the data is low. To maximise the amount of data when studying distributions
of fake jets, in the following plots data from entire fills are used while the
simulations correspond to the higher rate at the start of a fill. In order to compensate for this, the
MC results have been scaled by the ratio of "All Fill" to  "Fill Start"  values given in Table\,\ref{tabRateFJ}.

Figure\,\ref{fig:bananas} compares the distribution in the $\eta$--$\Delta t$ plane 
of fake jets seen in ATLAS data with the simulated rate of energy deposition clusters having \pt\!$>$16\,\GeV.
The pedestal, i.e.\ entries outside the banana area, seen in Figure\,\ref{fig:bananaDat} 
is mostly due to beam--gas and off-momentum halo background from ghost charge\,\cite{backgroundpaper2012}, 
a contribution that is not included in the simulations.\footnote{The data in the $\eta$--$\Delta t$ plot are restricted
to LHC fills prior to 3rd August when the LHC made chromaticity changes which caused a significant increase of
ghost charge\,\protect\cite{backgroundpaper2012}.}
A time-smearing due to the LHC bunch length has been applied to the simulated jet times. The instrumental
time resolution of the calorimeters depends on $\eta$ and the calorimeter cell energy. 
For Figure\,\ref{fig:bananaSim} 
the instrumental resolution was determined by fitting the width of the downstream tail of the banana shape in simulation, 
between $\eta=-3$ and $\eta=-2$, to the data. The fitted value of 0.5\,ns is consistent with the range of resolutions
measured for the ATLAS calorimeters.
The dashed horizontal lines indicate the jet-trigger time window of $\pm$12.5\,ns.  
Entries falling outside these lines are not seen in data, unless the event was selected by another trigger or has 
a sufficiently energetic subleading jet within the trigger window.
The position and curvature of the banana pattern within the trigger window is well reproduced by the simulations, 
which indicates that the jet times are correctly simulated.

The curvature of the banana shape depends on the radial position in the calorimeter: 
as indicated by Eq.\,(\ref{eq_jetTime}) and Figure\,\ref{fig:schematicFakeJetTime}, 
fake jets at $P_{2}$ will have a larger time advance 
than those at $P_{1}$ due to the difference in radial position. 
In Figure\,\ref{fig:bananas}, two banana shapes with slightly different curvature can be   
distinguished. The upper and lower tails correspond to fake jets in the LAr and Tile calorimeters
respectively.
At higher $|\eta|$ on the downstream side the bananas merge and the fake-jet times approach $\Delta t=0$,
although a small negative offset remains due to the dependence on $r$ in Eq.\,(\ref{eq_jetTime}).
Very early fake jets, with $\Delta t<-20$\,ns, are seen in the simulations. These are all in the 
upstream part of the barrel Tile calorimeter or the upstream endcaps. Falling outside the trigger window, they are not 
seen in the data. However, a minor concentration of jets at $\eta\!\approx\,$2 and $\Delta t\!\approx\,$10\,ns 
can be distinguished in Figure\,\ref{fig:bananaDat}. It is caused by upstream fake jets associated with the following 
bunch, which arrives 50\,ns later. These jets are reproduced by the simulations and are seen at the same $\eta$ 
but at $\Delta t\!\approx\,-40$\,ns in Figure\,\ref{fig:bananaSim}.

\begin{figure}[t]
\centering
\subfloat[]{\includegraphics[width=0.48\textwidth, trim=0cm 3.5cm 0cm 0cm, clip=true]{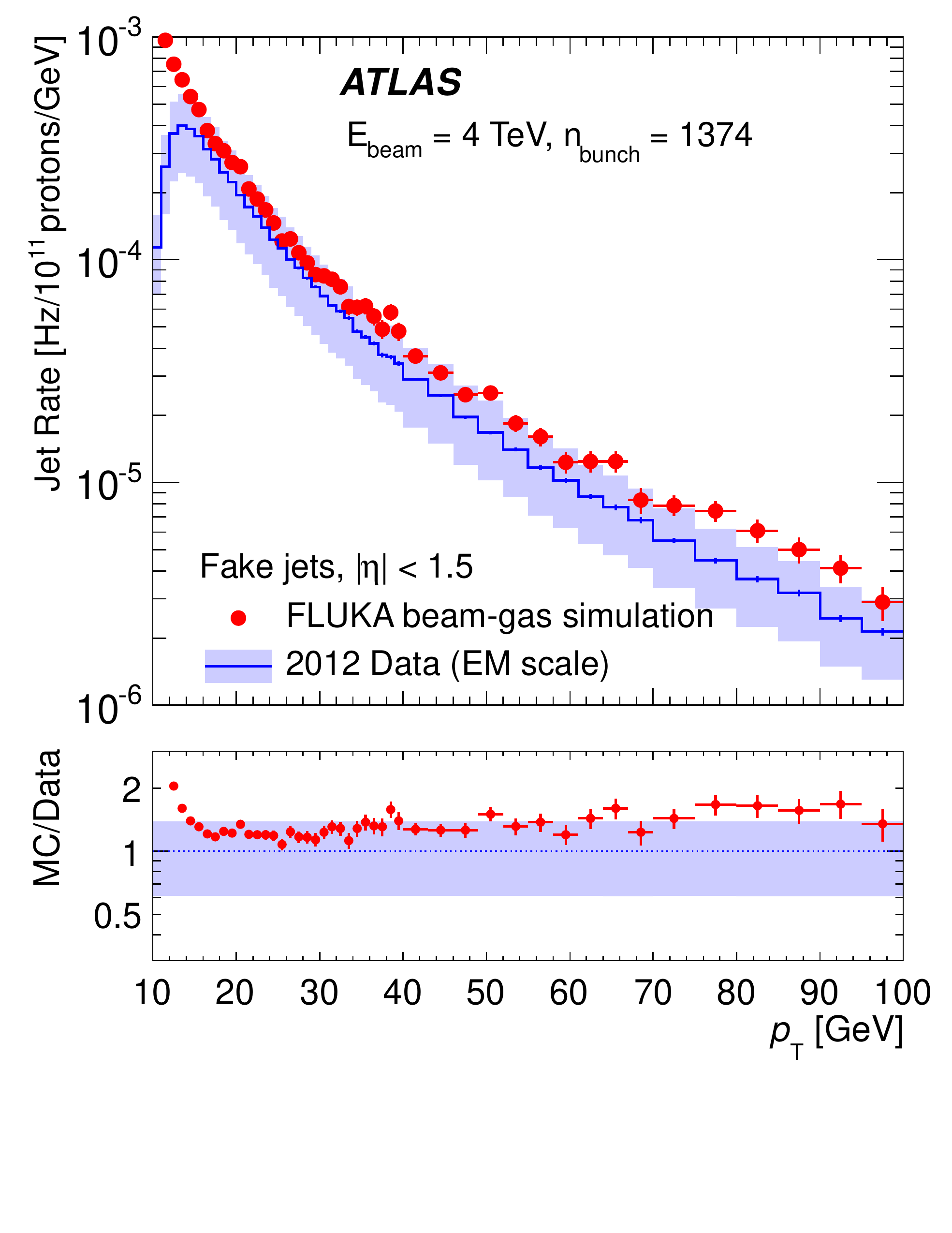}
\label{fig:fakeJetPtDistr}}
\subfloat[]{\includegraphics[width=0.48\textwidth, trim=0cm 3.5cm 0cm 0cm, clip=true]{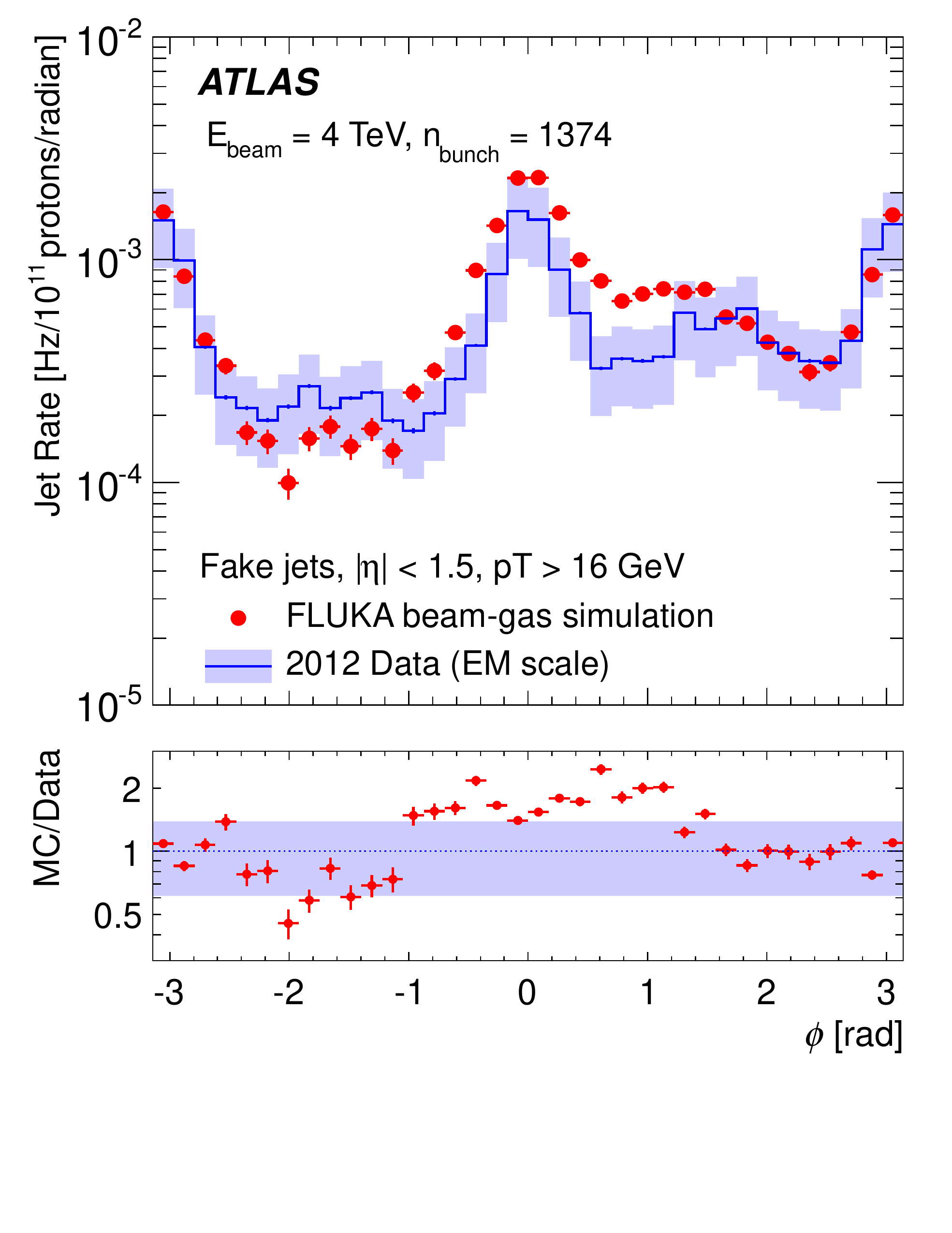}
\label{fig:fakeJetAzimuthal}}
\caption{
Distributions of (a) transverse momentum \pT and (b) azimuthal angle $\phi$ of fake jets with pseudorapidity $|\eta|<$1.5 
in data (blue histogram) compared with those of energy deposition clusters from \fluka beam--gas simulations in the same 
$\eta$ range (red circles). 
The \pt spectrum indicates that the ATLAS jet trigger reaches full efficiency only around 15\,\GeV. Therefore an additional 
requirement of \pt\!$>$16\,\GeV\ is applied to events in the azimuthal-angle distribution. The errors shown on the 
simulation results are statistical only. 
The blue band indicates the fill-to-fill variation of the data, while the small blue error bars show the uncertainty in 
the mean of all fills. Data from entire fills have been used in order to minimise statistical uncertainties. The simulations, which
correspond to the start-of-fill conditions have been scaled by the ratio of the total "All Fill" to "Fill Start" rates, 
shown in Table\,\ref{tabRateBCM}.
}
\label{fig:fakeJetComp}
\end{figure}

Figure\,\ref{fig:fakeJetComp} compares transverse-momentum and azimuthal-angle distributions of the simulated fake jets with data.
For these comparisons, jets within the banana area are extracted from data in order to minimise the contribution 
from the pedestal. 
As in the case of Figure\,\ref{fig:bcmTime}, the low count rates prevent an estimation of the fill-to-fill variation
in individual bins. The same approach as described for Figure\,\ref{fig:bcmTime} is adopted, i.e.\ the blue band 
shows the fill-to-fill variation determined from the total rates and the small error bars show the uncertainty in the
mean value in each bin.

The transverse-momentum distribution of the simulated fake jets, shown in Figure\,\ref{fig:fakeJetPtDistr},
continues on a rising slope below $\sim$15\,\GeV, while the data dip down. 
This is due to the jet trigger, used to select the data, which reaches full efficiency only above $\sim$15\,\GeV. 
When full trigger efficiency is reached, the simulations agree well with the data up to \pt$\approx$\,50\,\GeV, but towards 
higher transverse momenta the simulation tends to overestimate the data.

In Figure\,\ref{fig:fakeJetAzimuthal} an additional requirement of \pt$\!>$16\,\GeV\ is applied 
in order to select events above the trigger efficiency turn-on. The azimuthal distribution of fake jets from BIB
is well reproduced at a qualitative level.
The characteristic peaks at $\pm \pi$ and 0 are mainly due to the bending in the horizontal plane  
that occurs in the D1 and D2 dipoles and the LHC arc\,\cite{backgroundpaper2011}.
The lower rate at $-\pi/2$ compared to $\pi/2$ is due to the tunnel floor reducing the muon flux,  
as seen from the contours in Figure\,\ref{fig:muonsIntPlane}. A tendency of the simulation to overestimate the
data between $\phi=-1$ and $\phi=\pi/2$ and to underestimate around $\phi=-\pi/2$ is seen. A comparison with Figure\,\ref{fig:muonsIntPlane}  
suggests that these effects might be related to the tunnel geometry: the simulations underestimate at $-y$,
where the tunnel floor reduces the free drift space for pion and kaon decay, and overestimate around $+x$ where the 
horizontal offset provides extra space. Such differences could arise, for instance, if the $z$-distribution of 
the beam--gas events is not correct. A wrong $z$-distribution could change the impact of the reduced, or increased, 
free drift space. Another possibility is an inaccurate description of material around the beam line, which 
would affect the free drift space available. Forthcoming background measurements, with artificially introduced 
local pressure bumps, may shed some light on this.
Cosmic-ray muons, which can also produce fake jets, are included in the data but not in the simulation. Studies reported in 
Ref.\,\cite{backgroundpaper2012} indicate that the fake-jet rate at low \pt\ arising from cosmic-ray muons is less 
than 10\% of the total rate due to BIB. The radiative energy losses are point-like processes and since the flux of cosmic 
muons is uniform in space and time, the rate of fake jets produced by them should be independent of $\phi$.
However, due to the significant variation of the rate as a function of $\phi$, a visible 
contribution from cosmic-ray induced fake jets cannot be entirely excluded around $\phi=-\pi/2$, where the rates are lowest.

Although the differences in the shapes of the distributions shown in Figure\,\ref{fig:fakeJetComp} are not understood,
the agreement can be considered good, given the complexity of the entire simulation chain. The large systematic uncertainty
due to limited knowledge of the residual gas pressure distribution is the most likely cause of the differences seen,
although more detailed studies with localised and well-controlled pressure bumps will be needed to verify this.

\section{Origin of backgrounds}
\label{sect:zorigin}

\begin{figure}[t]
\centering
\includegraphics[width=0.98\textwidth, trim=0cm 4.0cm 0cm 0cm, clip=true]{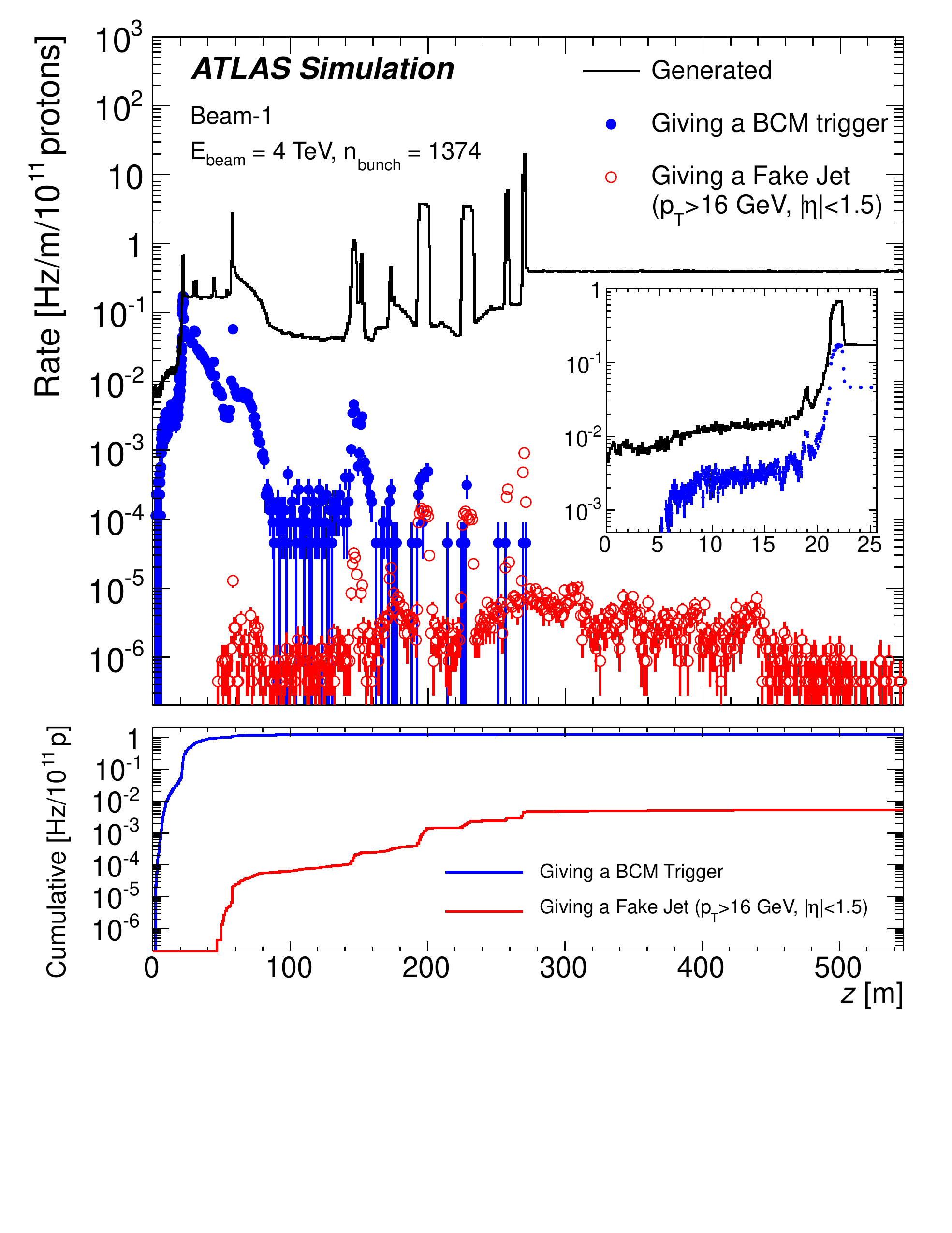}
\caption{
Distributions of the $z$-coordinate of the origin of simulated beam--gas events giving a BIB signature in ATLAS detectors. 
The solid blue circles show the $z$-distribution of events that give a BCM beam background signature while the 
open red circles show the $z$-distribution of beam--gas events that result in energy deposition clusters 
with \pt > 16\,\GeV\ and $|\eta|\!<$1.5 in the calorimeters. 
The $z$-distribution of the generated events (black line) corresponds to the beam--gas rate shown in 
Figure\,\protect\ref{fig:pint}, i.e.\ reflects the residual gas distribution in the beam pipe.
The small inset shows the region $z<25$\,m in more detail. The lower plot shows the cumulative rate, as a function 
of $z$, of events resulting in BCM background events or fake jets. At large $z$ these two histograms converge to the 
total simulated rates given in Tables\,\protect\ref{tabRateBCM} and \protect\ref{tabRateFJ} respectively.
}
\label{fig:zOrigin}
\end{figure}

Knowledge of the $z$-coordinate of the origin of each simulated event provides information beyond that which
can be extracted from the data. Figure\,\ref{fig:zOrigin} shows the distribution of the origins of the simulated events 
that give a BCM background trigger signature or generate a fake jet in the barrel calorimeters.
The black histogram, which shows the $z$-distribution of the generated events, reflects the residual gas 
distribution in the beam pipe and is equivalent to Figure\,\ref{fig:pint}. Most of the events with a BCM background trigger signature 
originate from the inner-triplet region ($z\!\approx\,$22--55\,m) 
with a small contribution from $z\!\approx\,$150\,m, where the tertiary collimator causes a local pressure bump.
This result is consistent with the observations made in the data, that the BCM background trigger rate correlates 
with the residual gas pressure measured by vacuum gauges at $z\!=\,$22\,m\,\cite{backgroundpaper2012}.
The fake jets, on the contrary, originate predominantly from more distant beam losses with pronounced spikes
at the locations of the 4.5\,K magnets. According to the simulations, about 10\% of fake-jet events are associated with beam--gas 
events in the LHC arc ($z>$270\,m), but this fraction depends strongly on the relative pressure in 4.5\,K and 1.9\,K sections.
The lower plot in Figure\,\ref{fig:zOrigin} shows the cumulative distributions corresponding to the histograms in the upper
plot. These highlight that practically all BCM background events originate from $z<60$\,m while only $\sim$1\% of fake-jet 
events are associated with beam--gas collisions in that $z$-range.
Since the simulations disfavour any significant correlation between BCM background and fake jets in the barrel calorimeters, they
suggest that BCM and fake-jet rates, seen in the data, can be used to disentangle backgrounds 
originating from different regions in $z$. 
As discussed before, the residual gas pressure depends on local properties of the beam pipe, such as material and temperature, and also on the 
radiation intensity. Therefore the pressure in different $z$-regions, as well as its uncertainties, can be considered to be uncorrelated. 
In particular, the prediction of BCM background, which originates predominantly 
from the inner triplet, depends on the accuracy of the pressure simulations within 1.9\,K magnets. Most of the
fake jets originate from the beam--gas events within 4.5\,K magnets where the desorption characteristics and, therefore, 
the gas composition, are  different. Thus it is not surprising to find, in Tables\,\ref{tabRateBCM} and \,\ref{tabRateFJ}, 
better agreement in one observable than in the other.

\FloatBarrier

\section{Conclusion}
Beam-induced background measurements in ATLAS during the 2012 LHC run with 4\,\TeV\ proton beams are compared with 
dedicated \fluka Monte Carlo simulations of the background due to inelastic beam--gas interactions.
Methods of extracting fake jets and BCM trigger signatures from a \fluka simulation
were developed and applied during simulations and the post-processing of the results, i.e.\
reconstruction of the background signatures.
The simulations, performed using a two-step method, agree within a factor of two with the rate of 
background trigger signatures in the BCM detector and the fake-jet rates observed in the ATLAS data.
This is well within the uncertainty in the residual gas pressure in the beam pipe.

Simulations reproduce rather well the shape of the time distribution of hits in the BCM as well as that of the fake jets 
in the pseudorapidity--time plane. The simulated spectrum of energy depositions in the calorimeters 
agrees with the spectrum of reconstructed transverse momenta of the observed fake jets although there is an
indication of an overestimate towards higher \pt. 
In the azimuthal distribution of the fake jets, the characteristic peaks in the horizontal plane are 
reproduced by the simulations, but differences are seen in some details of the structure in azimuthal angle. These might
be related either to inaccuracies in the pressure distribution or incomplete modelling of material close to the beam line.

The simulations indicate that background seen by the BCM originates mainly from the inner triplet region ($z<55$\,m)
while the majority of fake jets induced by beam--gas interactions have an origin at a distance of $z\!\gtrsim\,$150\,m from the interaction point.

The level of agreement between the simulations and measurement demonstrates the good understanding of beam background that has been
reached in the ATLAS experiment. It also illustrates the capability of the various simulation tools to reproduce the
beam background through a complex chain involving simulation of the residual gas pressure distribution, taking into
account various dynamic effects from the beam, transport of the beam--gas secondaries over 
long distances in the LHC magnet lattice and through the ATLAS detector, and 
finally the modelling of the reconstructed background signatures in ATLAS.

\section*{Acknowledgements}

\input{Acknowledgements}

\clearpage
\printbibliography

\clearpage

\include{atlas_authlist}

\end{document}

%% file: DAPR-2017-01-metadata.tex
\AtlasTitle{Comparison between simulated and observed LHC beam backgrounds 
in the ATLAS experiment at ${E_{\textrm {beam}}}\!=\!4$\,\TeV}

\author{The ATLAS Collaboration}

\AtlasAbstract{%
Results of dedicated Monte Carlo simulations of beam-induced background (BIB) in the ATLAS experiment at 
the Large Hadron Collider (LHC) are presented and compared with data 
recorded in 2012. During normal physics operation this background arises mainly from scattering 
of the 4\,\TeV\ protons on residual gas in the beam pipe.
Methods of reconstructing the BIB signals in the ATLAS detector, developed and implemented in the simulation 
chain based on the \textsc{Fluka}\xspace Monte Carlo simulation package, are described.
The interaction rates are determined from the residual gas pressure distribution in the LHC ring 
in order to set an absolute scale on the predicted rates of BIB so that they can be compared quantitatively 
with data. Through these comparisons the origins of the BIB leading to different observables in the ATLAS 
detectors are analysed. 
The level of agreement between simulation results and BIB measurements by ATLAS in 2012
demonstrates that a good understanding of the origin of BIB has been reached.
}

\AtlasRefCode{DAPR-2017-01}

\PreprintIdNumber{CERN-EP-2018-240}

\AtlasDate{\today}

\arXivId{1810.04450}

\AtlasJournalRef{JINST 13 (2018) P12006}
\AtlasDOI{10.1088/1748-0221/13/12/P12006}

%% file: Acknowledgements.tex

We thank CERN for the very successful operation of the LHC, as well as the
support staff from our institutions without whom ATLAS could not be
operated efficiently.
The residual pressure map used in this study is courtesy of C. Yin Vallgren.

We acknowledge the support of ANPCyT, Argentina; YerPhI, Armenia; ARC, Australia; BMWFW and FWF, Austria; ANAS, Azerbaijan; SSTC, Belarus; CNPq and FAPESP, Brazil; NSERC, NRC and CFI, Canada; CERN; CONICYT, Chile; CAS, MOST and NSFC, China; COLCIENCIAS, Colombia; MSMT CR, MPO CR and VSC CR, Czech Republic; DNRF and DNSRC, Denmark; IN2P3-CNRS, CEA-DRF/IRFU, France; SRNSFG, Georgia; BMBF, HGF, and MPG, Germany; GSRT, Greece; RGC, Hong Kong SAR, China; ISF and Benoziyo Center, Israel; INFN, Italy; MEXT and JSPS, Japan; CNRST, Morocco; NWO, Netherlands; RCN, Norway; MNiSW and NCN, Poland; FCT, Portugal; MNE/IFA, Romania; MES of Russia and NRC KI, Russian Federation; JINR; MESTD, Serbia; MSSR, Slovakia; ARRS and MIZ\v{S}, Slovenia; DST/NRF, South Africa; MINECO, Spain; SRC and Wallenberg Foundation, Sweden; SERI, SNSF and Cantons of Bern and Geneva, Switzerland; MOST, Taiwan; TAEK, Turkey; STFC, United Kingdom; DOE and NSF, United States of America. In addition, individual groups and members have received support from BCKDF, CANARIE, CRC and Compute Canada, Canada; COST, ERC, ERDF, Horizon 2020, and Marie Sk{\l}odowska-Curie Actions, European Union; Investissements d' Avenir Labex and Idex, ANR, France; DFG and AvH Foundation, Germany; Herakleitos, Thales and Aristeia programmes co-financed by EU-ESF and the Greek NSRF, Greece; BSF-NSF and GIF, Israel; CERCA Programme Generalitat de Catalunya, Spain; The Royal Society and Leverhulme Trust, United Kingdom. 

The crucial computing support from all WLCG partners is acknowledged gratefully, in particular from CERN, the ATLAS Tier-1 facilities at TRIUMF (Canada), NDGF (Denmark, Norway, Sweden), CC-IN2P3 (France), KIT/GridKA (Germany), INFN-CNAF (Italy), NL-T1 (Netherlands), PIC (Spain), ASGC (Taiwan), RAL (UK) and BNL (USA), the Tier-2 facilities worldwide and large non-WLCG resource providers. Major contributors of computing resources are listed in Ref.~\cite{ATL-GEN-PUB-2016-002}.

%% file: atlas_authlist.tex

\begin{flushleft}
{\Large The ATLAS Collaboration}

\bigskip

M.~Aaboud$^\textrm{\scriptsize 34d}$,
G.~Aad$^\textrm{\scriptsize 99}$,
B.~Abbott$^\textrm{\scriptsize 124}$,
O.~Abdinov$^\textrm{\scriptsize 13,*}$,
B.~Abeloos$^\textrm{\scriptsize 128}$,
S.H.~Abidi$^\textrm{\scriptsize 165}$,
O.S.~AbouZeid$^\textrm{\scriptsize 143}$,
N.L.~Abraham$^\textrm{\scriptsize 153}$,
H.~Abramowicz$^\textrm{\scriptsize 159}$,
H.~Abreu$^\textrm{\scriptsize 158}$
% ...
\bigskip
\\

% ...
$^{13}$Institute of Physics, Azerbaijan Academy of Sciences, Baku; Azerbaijan.\\
% ...
$^{34}$$^{(a)}$Facult\'e des Sciences Ain Chock, R\'eseau Universitaire de Physique des Hautes Energies - Universit\'e Hassan II, Casablanca;$^{(b)}$Centre National de l'Energie des Sciences Techniques Nucleaires (CNESTEN), Rabat;$^{(c)}$Facult\'e des Sciences Semlalia, Universit\'e Cadi Ayyad, LPHEA-Marrakech;$^{(d)}$Facult\'e des Sciences, Universit\'e Mohamed Premier and LPTPM, Oujda;$^{(e)}$Facult\'e des sciences, Universit\'e Mohammed V, Rabat; Morocco.\\
% ...
$^{99}$CPPM, Aix-Marseille Universit\'e, CNRS/IN2P3, Marseille; France.\\
% ...
$^{124}$Homer L. Dodge Department of Physics and Astronomy, University of Oklahoma, Norman OK; United States of America.\\
% ...
$^{128}$LAL, Universit\'e Paris-Sud, CNRS/IN2P3, Universit\'e Paris-Saclay, Orsay; France.\\
% ...
$^{143}$Santa Cruz Institute for Particle Physics, University of California Santa Cruz, Santa Cruz CA; United States of America.\\
% ...
$^{153}$Department of Physics and Astronomy, University of Sussex, Brighton; United Kingdom.\\
% ...
$^{158}$Department of Physics, Technion, Israel Institute of Technology, Haifa; Israel.\\
$^{159}$Raymond and Beverly Sackler School of Physics and Astronomy, Tel Aviv University, Tel Aviv; Israel.\\
% ...
$^{165}$Department of Physics, University of Toronto, Toronto ON; Canada.\\
% ...
$^{a}$ Also at  Department of Physics, University of Malaya, Kuala Lumpur; Malaysia.\\
$^{b}$ Also at Borough of Manhattan Community College, City University of New York, NY; United States of America.\\
$^{c}$ Also at California State University, East Bay; United States of America.\\
$^{d}$ Also at Centre for High Performance Computing, CSIR Campus, Rosebank, Cape Town; South Africa.\\
$^{e}$ Also at CERN, Geneva; Switzerland.\\
% ...
$^{*}$ Deceased

\end{flushleft}